\documentclass[12pt,a4paper]{article}
\usepackage{caption}
\usepackage{graphicx} 

\usepackage{appendix}
\usepackage{caption}
\usepackage{subcaption}
\usepackage{makecell}

\usepackage[backend=biber,style=nature]{biblatex}

\usepackage{color}

\usepackage{amsmath,amssymb,amsthm}
\usepackage{hyperref}
\usepackage[margin=1in,letterpaper,top=100pt,bottom=100pt,left=68pt,right=66pt]{geometry} 
\usepackage{xcolor}
\usepackage{setspace}
\usepackage{float}
\usepackage{csquotes}
\usepackage{authblk}
\usepackage{booktabs}

\newcommand{\beginsupplement}{%
        \setcounter{table}{0}
        \renewcommand{\thetable}{S\arabic{table}}%
        \setcounter{figure}{0}
        \renewcommand{\thefigure}{S\arabic{figure}}%
        \renewcommand{\thesection}{S}
     }

\title{Local irreversible phase reconfiguration and thermal-memory effects in a highly-correlated manganite.} 

\author[1]{G. Kuhl-Soares}
\author[1]{O. Canton}
\author[1]{E. Granado*}
\author[2]{D. Carranza-Célis}
\author[1]{M. Knobel}
\author[1]{G. B. Gomide}
\author[2]{J. G. Ramirez*}
\author[1]{D. Muraca*}

\affil[1]{\small Universidade Estadual de Campinas (UNICAMP), Instituto de F\'isica Gleb Wataghin (IFGW), Campinas 13083-859, Brazil}
\affil[2]{\small Department of Physics, Universidad de los Andes, Bogotá 111711, Colombia}
\date{}
\begin{document}

\maketitle

\begin{abstract}
    Phase-separated manganites provide a unique platform to study the dynamics of competing electronic and structural orders in correlated systems. In La$_{0.275}$Pr$_{0.35}$Ca$_{0.375}$MnO$_3$ (LPCMO), we use temperature-cycling Raman spectroscopy to uncover a previously unidentified regime of structural irreversibility, emerging from the interplay between lattice distortions and phase competition across the phase-separation and charge–orbital ordering temperatures. This irreversible behavior encodes a thermal-memory effect reflecting the system’s history-dependent energy landscape. Correlated magnetic and transport responses confirm the coupling between lattice and electronic degrees of freedom, revealing a new form of nonequilibrium phase dynamics in mixed-valence oxides. These results advance the understanding of metastability and memory phenomena in strongly correlated materials. 
\end{abstract}

Mixed-valence manganites exemplify the complexity inherent to strongly correlated oxides, wherein the interaction among spin, charge, orbital, and lattice degrees of freedom gives rise to emergent properties with tunable functionalities. The phase-separation (PS) regime intrinsic to these materials, characterized by the coexistence of ferromagnetic metallic (FMM) and antiferromagnetic insulating with charge and orbital ordering (AFMI-COO) regions, imparts a notable sensitivity to external perturbations. Modest magnetic, electric, or thermal stimuli can dynamically reconfigure the PS landscape, producing resistive-switching and memory effects that have been widely explored for neuromorphic and adaptive electronics \cite{Sacanell2004, Sacanell2018, Levy2002, Quintero2007, Gomide2025, Ivan2024, Schulman2024, Paasonen2024, Jaman2025, Zhang2020, Lahteenlahti2021,li2022,Zhou2015,Milloch2024}.  

Among the extensively studied manganites, La$_{5/8-y}$Pr$_y$Ca$_{3/8}$MnO$_3$ (LPCMO) serves as a prototypical system to investigate PS phenomena. At intermediate Pr concentrations, LPCMO exhibits sub-micrometric coexistence of FMM and AFMI-COO phases over a wide temperature range, giving rise to a dynamic-like phase separation (DPS) regime between approximately 30 and 100 K \cite{Sarma2004, Uehara, Ghivelder2005, Garcia2011, Dhakal2007, Diego24}. In this regime, unpinned interfacial walls fluctuate and respond readily to magnetic and electric fields \cite{Uehara, Dhakal2007, Merten, Ghivelder2005}, mechanical strain \cite{Podzorov2001, Ahn2004, Ogimoto2005,Ward2009, Kim2021}, pressure \cite{Laukhin1997, Baldini2012}, light \cite{Deng2024, Lin2018}, and the compound's thermal history \cite{kundhikanjana2015,Sarma2004, Sacanell2004, Ward2011, Sacanell2018}.  

Beyond contrasting electronic and magnetic behavior, the two coexisting phases also differ structurally. While the FMM regions maintain the compound's room-temperature orthorhombic \textit{Pnma} symmetry, the AFMI-COO matrix exhibits a monoclinic \textit{P2$_1$/m} structure, originating from cooperative Jahn-Teller (JT) distortions of Mn$^{3+}$ ions and associated charge and orbital ordering \cite{Collado2003, Radaelli1997, Goff2004, Garcia2011}. The contrast between long- and short-range JT distortions provides a direct structural fingerprint of the PS evolution. Raman spectroscopy, noted for its exceptional sensitivity to lattice distortions and orbital-phonon coupling, is particularly adept at probing these effects in real time \cite{Granado1998, Granado2000, Kim, Merten, Dediu2000, Abrashev1999, Adams2000, Moshnyaga2014, Granado2001}.

Laser-based approaches extend the control landscape of correlated systems by directly engaging the fundamental lattice, spin, and orbital degrees of freedom that define their collective behavior. Ultrafast optical excitation can coherently drive phonons, modulate exchange interactions, and reshape electronic bandwidths on femtosecond timescales. Unlike electrical or magnetic stimuli, which act non-locally through current flow or field gradients, laser perturbations are intrinsically local and contactless, enabling precise spatial control with negligible dissipation. This selectivity grants access to hidden metastable configurations and non-equilibrium pathways unattainable under static driving, leading to light-induced resistive and magnetic switching \cite{Deng2024, Lin2018}. Such all-optical schemes offer a powerful route toward \textit{quantum} and \textit{neuromorphic devices}, where coherence preservation, minimal Joule heating, and nanoscale phase engineering are critical for scalable operation \cite{Kalcheim2020}.

Here, we employ temperature-cycling Raman spectroscopy to uncover a previously unreported regime of structural irreversibility associated with the reconfiguration of the DPS network in LPCMO. By applying cumulative, pulsed-like temperature variations between low base, T$_B$, and target, T$_T$, temperatures, we observe metastable lattice states imprinted in the intensity and width of the JT stretching modes. Complementary magnetic and transport measurements corroborate that these structural modifications coincide with the reorganization of the PS network. Altogether, these findings demonstrate that targeted thermal and optical protocols can be used to manipulate the balance between coexisting FMM and AFMI-COO phases at will. This establishes a pathway for laser-assisted phase engineering in correlated oxides. 

Polycrystalline LPCMO ($y=0.35$) samples were prepared using a standard solid-state reaction method, as described by \cite{Collado2003,li2008synthesis, Diego24}. Its orthorhombic Pnma (a $\approx$ c $\approx$ $\sqrt{2}$ b) single phase character at room temperature was confirmed by synchrotron x-ray powder diffraction performed in the PAINEIRA beamline in CNPEM-SIRIUS \cite{PAINEIRA, Collado2003}. The Rietveld refinement performed with the GSAS-II software \cite{GSAS} and the refined crystallographic parameters can be found in figure \ref{fig.XRD.ref} and table \ref{tab.XRD} in Supplementary Materials \ref{Supplementary Materials}. 

\begin{figure}[htbp!]
    \centering
    \includegraphics[width=.75\linewidth]{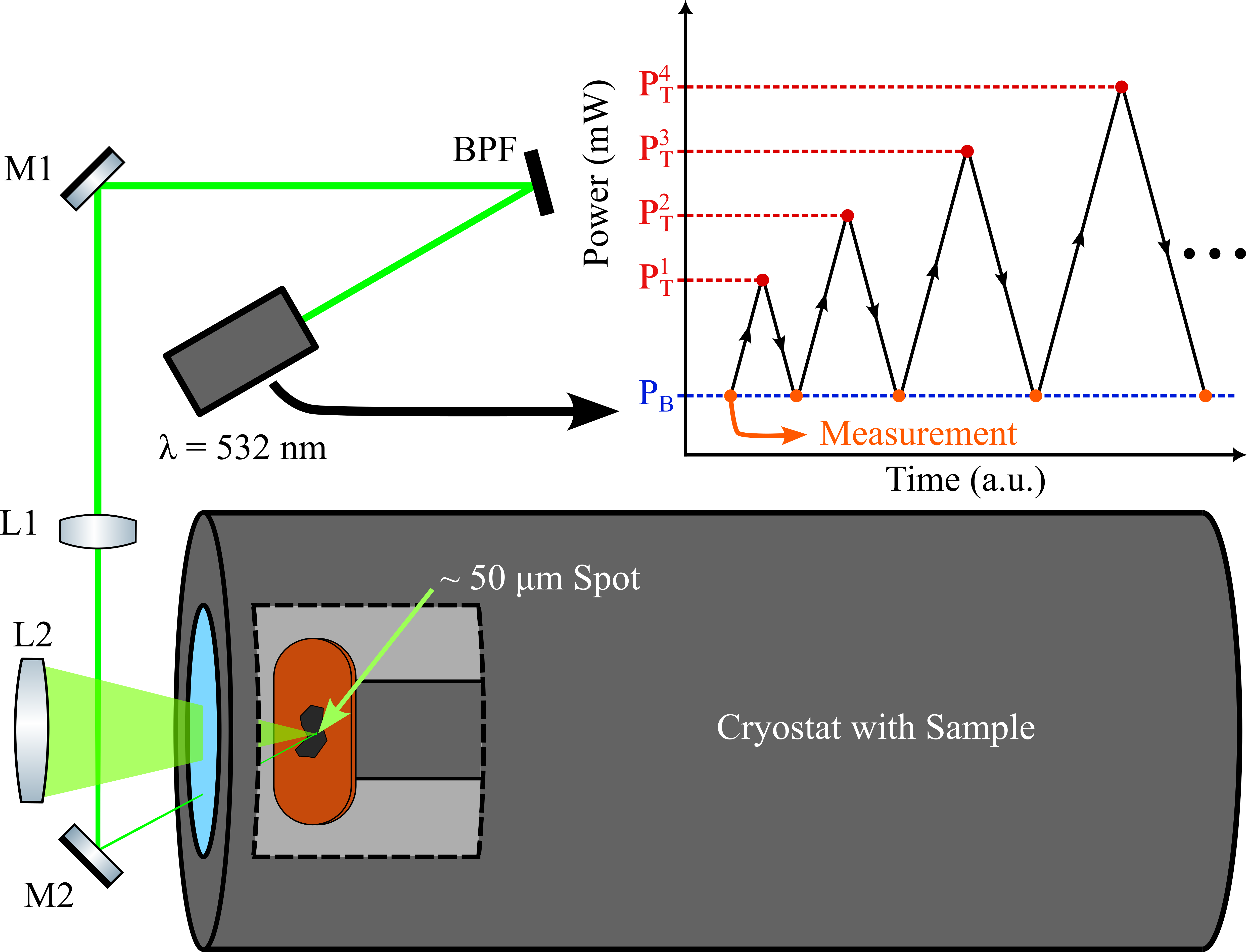}
    \caption{Part of the optical setup used for the Raman spectroscopy measurements together with the adopted experimental procedure, consisting of cyclic variations of laser power between a base, P$_B$, and target, P$_T^i$, values.}
    \label{fig.Panel_1}
\end{figure}

Raman scattering experiments were performed in a quasi-backscattering geometry using a 532 nm laser line with a spot size at the sample of $\approx$ 50 $\mu$m, as depicted in figure \ref{fig.Panel_1}. A Jobin-Yvon T64000 spectrometer with a single 600 mm$^{-1}$ grating and a set of Bragg notch filters was used. A LN$_2$-cooled multichannel charge-coupled device (Symphony II, Horiba) collected the scattered data. The sample was glued on the cold finger of a closed-cycle He cryostat with a minimal temperature of $\sim 15$ K. Spectra as a function of the cold-finger nominal temperature were collected while heating to investigate the thermal evolution of the sample's vibrational modes and to serve as a control parameter for the measurements performed as a function of laser power. This protocol will be referred to as "standard procedure/heating". In the measurements as a function of laser power, the sample was first cooled and maintained at the cryostat minimal temperature. Then, two different protocols were employed: 1) a continuous increase in power, with spectra collection using 10 mW $\rightarrow$ to 150 mW (with 10 mW steps), and 2) cycling the power, where spectra were collected at 10 mW (base power) before and after it was increased to target powers ranging from 20 mW to 150 mW, illustrated in figure \ref{fig.Panel_1}. The latter method of temperature variation allows for almost instantaneous heating and subsequent cooling of the analyzed region, and will be referred as "cyclic procedure/heating". The comparison between the spectra collected in protocol 1) and those collected in the standard procedure yielded the temperature of the analyzed spot as a function of power, as detailed in the Supplementary Materials \ref{Calibration}. The thermal cycles were conducted between T$_B$ = 38(4) K and T$_T$ = 59(4), 81(4), ..., 317 (7) K; see table \ref{tab.1} in the Supplementary Materials \ref{Calibration}. 
In both experiments, the laser spot probed the same position on the sample to enhance control over the localized heating effects induced by the laser, and the collection time was proportionally decreased as the laser power increased. Different spots and pieces of the sample were measured with the same protocol.

Magnetic and electric characterizations were performed using a MPMS 3 and PPMS Model 6000 (Quantum Design), respectivelly. Magnetization as a function of temperature, M-T, was measured in the VSM mode with an applied field of 50 Oe. Extractions were performed between 300 K and 2 K, cooling and warming at a rate of 2 K/min. The sample resistance was measured using a standard two-probe technique with spacing between analyzer probes of $\sim$ 0.7 mm and a compliance current of 0.1 $\mu$A. Resistance as a function of temperature, R-T, was measured between 300 K and 10 K, cooling and heating at a rate of 2 K/min, respectively. Magnetization and resistance as a function of the cyclic procedure between a base, T$_B$ = 30 K, and target, T$_T$, 50, 60, 70, ... 250 K, temperatures were measured after cooling to 30 K. Warming and cooling M-T and R-T curves between T$_B$ and T$_T$ were collected with a cooling/warming rate of 5 K/min.

\begin{figure}[htbp!]
    \centering
    \includegraphics[width=.75\linewidth]{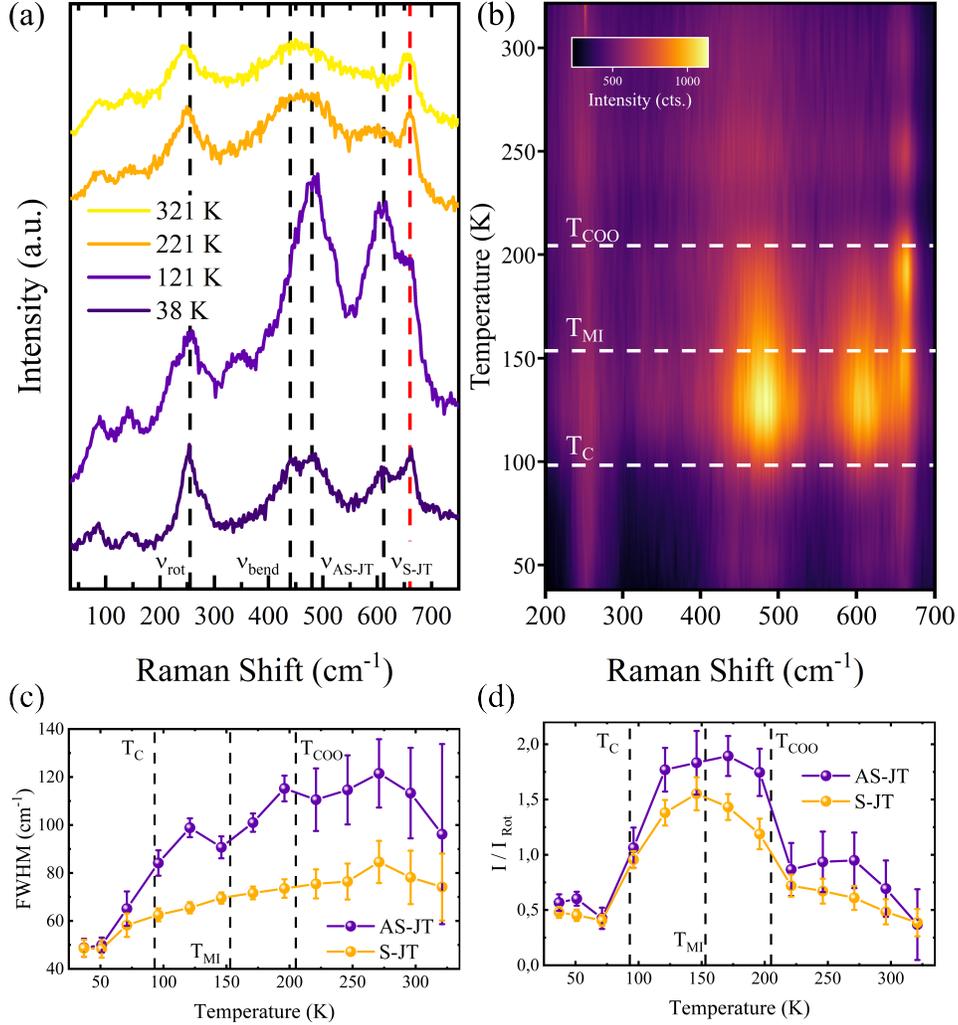}
    \caption{Figure (a) show the LPCMO Raman spectra for 38, 121, 221 and 321 K, vertically translated. The position of the main rotational ($\nu_{rot}$), bending ($\nu_{bend}$), AS-JT ($\nu_{AS-JT}$) and S-JT ($\nu_{S-JT})$ modes are indicated by black dashed vertical lines. In figure (b), all spectra are interpolated to produce the displayed heat map. The horizontal white lines indicates the sample's transitions temperatures. Figures (c) and (d) shows the FWHM and the normalized intensities of the AS- and S-JT modes. The intensities of the modes at each temperature were normalized by the $\sim$ 251 cm$^{-1}$ rotational mode intensity. In these figures, the black vertical dashed lines indicates the transition temperatures of the sample. All measurements were taken with fixed laser power ($P=10$ mW).}
    \label{fig.Panel_2}
\end{figure}

Figure \hyperref[fig.Panel_2]{\ref*{fig.Panel_2}} presents the standard thermal (cold finger) evolution of the LPCMO Raman spectrum. Figure \hyperref[fig.Panel_2]{\ref*{fig.Panel_2} (a)} shows spectra collected at T = 38, 121, 221 and 321 K. The position of the main rotational ($\nu_{rot} \sim$ 251 cm$^{-1}$), bending ($\nu_{bend} \sim$ 440 cm$^{-1}$), asymmetric, AS-JT ($\nu_{AS-JT}$ $\sim$ 485 cm$^{-1}$) and symmetric, S-JT ($\nu_{S-JT}$ $\sim$ 615 cm$^{-1}$), distortion modes \cite{Carron, Granado2000, Kim, Iliev1998, Merten, Iliev2003, Granado1998, Liarokapis1999,Merten2019, Amelitchev2001} are indicated by black dashed vertical lines. The red dashed vertical line indicates the presence of a spurious peak, possibly due to local oxygen vacancies. Figure \hyperref[fig.Panel_2]{\ref*{fig.Panel_2} (b)} displays an intensity map of the Raman spectra, with indications of the transition temperatures of the sample represented by dashed horizontal lines. These spectra show the AS-JT and JT modes intensities increasing significantly around 100 K, which is accompanied by the appearance of extra rotational and bending modes around 220, 330 and 515 cm$^{-1}$ \cite{Kim, Merten}, respectively. Figures \hyperref[fig.Panel_2]{\ref*{fig.Panel_2} (c)} and \hyperref[fig.Panel_2]{\ref*{fig.Panel_2} (d)} display the full widths at half maximum (FWHM) and intensities of the AS-JT and S-JT, respectively. The intensities were normalized by the rotational mode present at $\sim$ 250 cm$^{-1}$, which was taken as a reference. These modes broaden and become stronger as the system evolves from the DPS to the AFMI-COO regimes (T$_C$). The maximum intensities of the JT modes occur around the metal-insulator transition temperature, T$_{MI}$. Above T$_{COO}$, the charge and orbital ordering temperature, these modes weaken substantially.

\begin{figure}[htbp!]
    \centering
    \includegraphics[width=.65\linewidth]{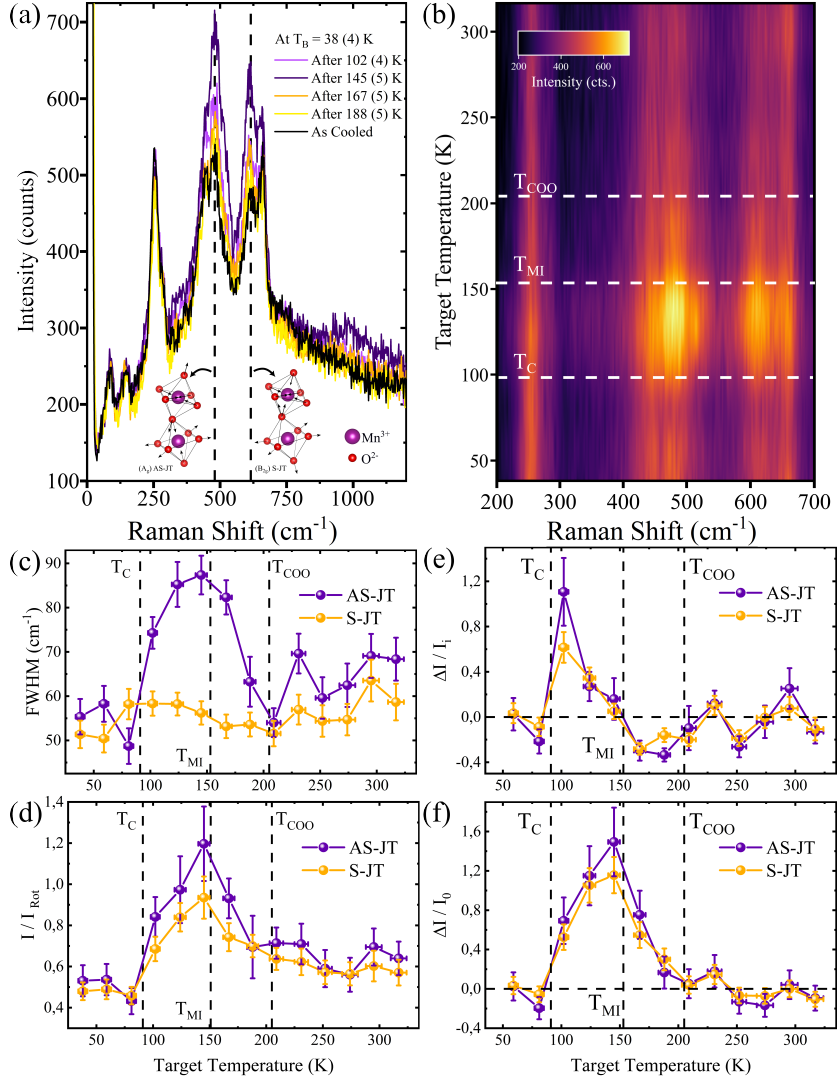}
    \caption{Figure (a) shows the LPCMO Raman spectra at T$_B$ before the thermal cycling procedure, labeled "as-cooled", and after specific cycling temperatures. Figure (b) shows a heat map produced by the interpolation of the Raman spectra at T$_B$ as a function of T$_T$, with the white horizontal dashed lines indicating the sample's transitions temperatures. Figures (c) and (d) show the FWHM and the normalized intensities of the AS- and S-JT modes as a function of T$_T$. The intensities of the modes at T$_B$ after each cycle were normalized by the corresponding $\sim$ 250 cm$^{-1}$ rotational mode intensity. Figures (e) and (f) display the normalized difference in intensity, $(I_F - I)/ I$, in both modes. The difference between each individual cycle, with $I=I_i$ the initial intensity before each cycle is shown in (e). While (f) shows the normalized difference between the final intensities values after each cycle and the initial, as-cooled, intensity, labeled $I =I_0$. In figures (c) - (f) the black vertical dashed lines correspond to the sample's transitions temperatures.}
    \label{fig.Panel_2.2}
\end{figure}

Figure \hyperref[fig.Panel_2.2]{\ref*{fig.Panel_2.2} (a)} displays the ``as-cooled" spectrum, collected before starting the thermal cycling procedure with laser power, in black, as well as spectra collected at T$_B$ after after successive thermal cycles between T$_B$ and T$_T$ =  102(4), 124(5), 167 (5) and 188 (5) K. Is revealed that both asymmetric (AS) and symmetric (S) JT distortion modes gain intensity relative to the others until 145 (5) K. For cycles with higher T$_T$, the intensities of these modes decrease continuously up to T$_T$ = 188(5) K. For this and higher values of T$_T$ (not shown), the Raman spectra after cycling are very similar to the ``as-cooled" one. These irreversible changes occur only within a specific temperature window bounded by the DPS critical (T$_C$) and charge-orbital ordering (T$_{COO}$) temperatures and vanish for T $>$ T$_{COO}$, revealing a distinct thermal-memory effect in the lattice degree of freedom.

Figure \hyperref[fig.Panel_2.2]{\ref*{fig.Panel_2.2} (b)} shows a color map of the Raman intensities at $T_B = 38\,\mathrm{K}$ as a function of the laser-induced local target temperature $T_T$. The similarity of this map with the one shown in figure \hyperref[fig.Panel_2]{\ref*{fig.Panel_2} (b)} is remarkable, despite the fact that these are completely different experimental protocols. The JT modes start to gain intensity around T$_T$ = T$_{C}$, with maximum intensity close to T$_T$ = T$_{MI}$, and then lose intensity as T$_T$ approaches T$_{COO}$ ($\sim$ 205 K). From T$_T$ = T$_{COO}$ onward, the mode intensities return to their original values, \textit{i.e.}, they are reversed or reset. Figure \hyperref[fig.Panel_2.2]{\ref*{fig.Panel_2.2} (c)} shows the FWHM of the JT modes as a function of T$_T$. As T$_T$ increases, the AS-JT mode undergoes a broadening followed by a sharpening, while the width of the S-JT mode shows less significant variation. This, in conjunction with the normalized mode intensities in figure \hyperref[fig.Panel_2.2]{\ref*{fig.Panel_2.2} (d)}, reveals that the thermal cycling process induces an enhancement, between T$_C$ $<$ T$_T$ $<$ T$_{COO}$, with a maximum around T$_{MI}$, followed by a weakening of COO phase at T$_B$. That is, the strength of the COO phase achieved at T$_T$ is retained when the system cools back to T$_B$.

The normalized intensities of both modes, in both protocols, figures \hyperref[fig.Panel_2]{\ref*{fig.Panel_2} (d)} and \hyperref[fig.Panel_2.2]{\ref*{fig.Panel_2.2} (d)}, show a quantitative correspondence for T$_T$ $\&$ T $<$ T$_{C}$. However, for T$_{C}$ $<$ T$_T$ $\&$ T  $<$ T$_{COO}$, such an agreement is not observed (apart from T$_T$ $\&$ T  $\sim$ 100 K in the AS-JT mode). Since what is being shown is the relative intensity between two distinct vibrational modes, this can be interpreted as a partial loss of information during the cooling process. The state reached at T$_B$ after thermal cycling for temperatures between T$_{C}$ $<$ T$_T$ $<$ T$_{COO}$ does not fully recover the state to which it was taken due to heating. Hence, during the cooling process, and due to the high dynamism of the phases throughout the temperature range considered here \cite{Ghivelder2005, Dhakal2007, Lee2002}, a new rearrangement of the phases occurs. However, this is not sufficient to completely "erase" the information stored by the material due to the thermal process to which it has been exposed.

It is remarkable how, when performing the thermal cycling process at temperatures T$_{C}$ and T$_{COO}$, the system has the capacity to partially \textit{recover} and \textit{reset} the state it reached at the different target temperatures. With the process of cyclic temperature variation, characteristics of the vibrational modes found in a conventional heating process are reached and stored at the base temperature of the system. This capacity to store and erase seems to be closely related to the material's well-known transitions \cite{Ghivelder2005, Dhakal2007, Lee2002, Kiryukhin2000, Diego24}. Performing cumulative thermal cycles that transition from its DPS to its mostly AFMI-COO regime, T$_C$, produces an intensity gain in its JT distortion modes that is similar to the gain observed when the compound is heated conventionally. Cyclically varying the temperature to T$_T$ $\geq$ T$_{COO}$ has the effect of erasing this intensity gain and reverting the system to its initial state.

In order to quantify these intensity changes, figures \hyperref[fig.Panel_2.2]{\ref*{fig.Panel_2.2} (e) and (f)} show the normalized difference, $(I_F - I)/ I$, between the initial $I$ and final intensity values, $I_F$, considering each individual cycle (figure \hyperref[fig.Panel_2.2]{\ref*{fig.Panel_2.2} (e)}) with $I=I_i$, and the initial "as-cooled" intensity values $I=I_0$ (figure \hyperref[fig.Panel_2.2]{\ref*{fig.Panel_2.2} (f)}). The individual contribution to the modes intensities of each thermal cycle reaches its maximum between T$_C$ $<$ T$_T$ $<$ T$_{MI}$. The intensity relation is reversed for T$_{MI}$ $<$ T$_T$ $<$ T$_{COO}$, \textit{i.e.}, $I_f$ starts to diminish in relation to I$_i$. For T$_T$ $>$ T $\geq$ T$_{COO}$, within the fitting error, no further changes are observed due to thermal cycling. In figure \hyperref[fig.Panel_2]{\ref*{fig.Panel_2} (f)} , it can be seen how, with respect to the initial "as-cooled" JT mode intensity values, an almost two-fold gain is observed between T$_C$ and T$_{MI}$. This gain diminishes as the target temperature approaches T$_{COO}$, and finally, it is reset after this transition, returning to values close to the initial ones.

\begin{figure}[htbp!]
    \centering
    \includegraphics[width=.75\linewidth]{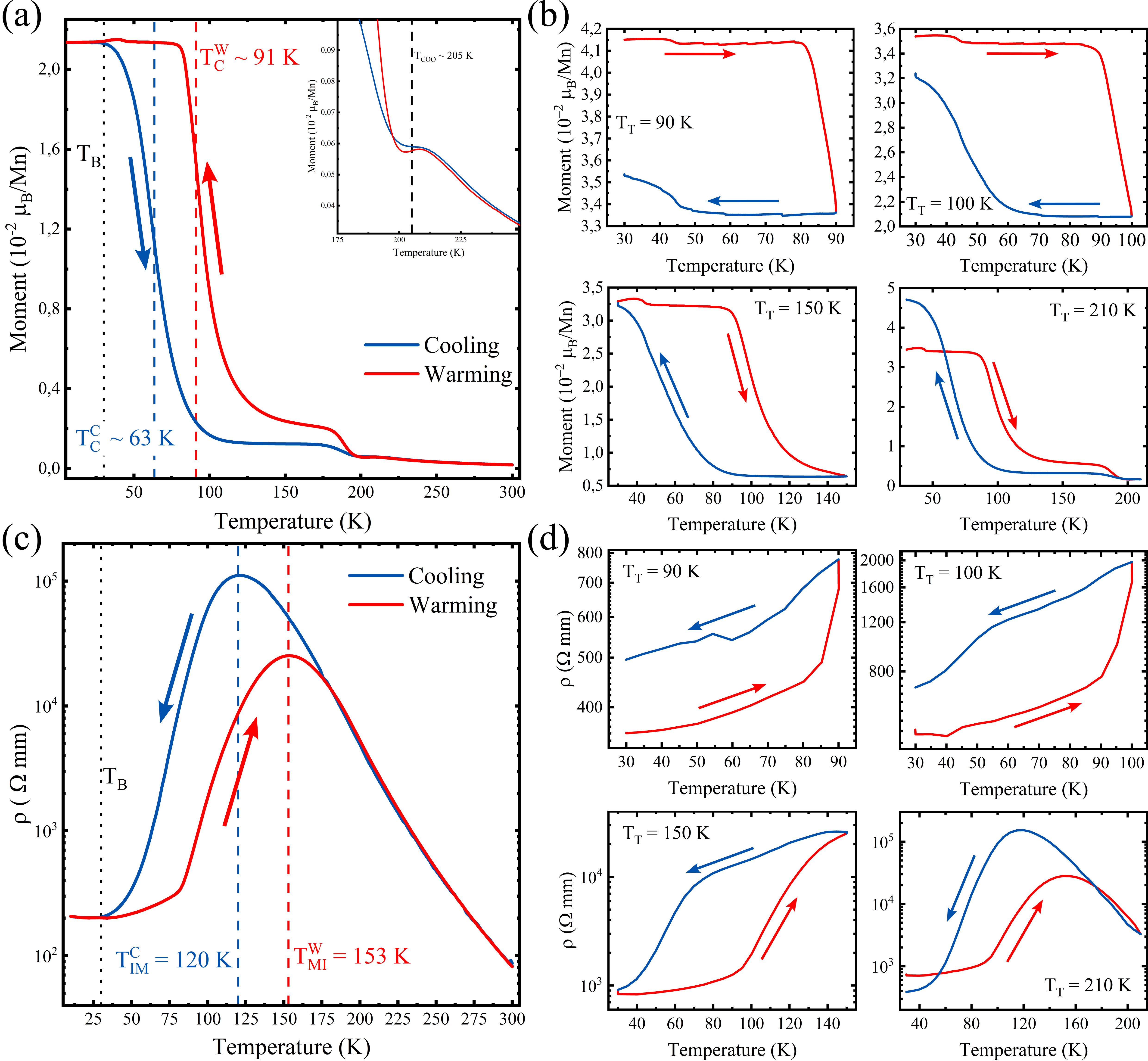}
    \caption{Figures (a) and (b) shows magnetization curves as a function of temperature, in units of $\mu_B$/Mn, in a cooling and heating regimes between 300 and 10 K and between T$_B$ = 30 K and T$_T$ = 90, 100, 150, and 210 K, respectively. Figures (c) and (d), similarly, shows resistivity curves as a function of temperature, in units of $\Omega$ mm, in a cooling and heating regime between 300 and 10 K and between T$_B$ = 30 K and T$_T$ =  90, 100, 150, and 210 K. }
    \label{fig.Panel_3}
\end{figure}

Complementary magnetization and resistivity measurements as a function of analogous thermal cycles were performed to correlate local Raman results with transport and magnetic macroscopic properties. Figures \hyperref[fig.Panel_3]{\ref*{fig.Panel_3} (a) and (b)} shows the full "Cooling"/"Warming" magnetization curves as a function of temperature (M - T), collected by cooling from 300 to 10 K and heating from 10 to 300 K, together with different M - T curves between T$_B$ and the values of interest of T$_T$, respectively. In the full M - T curve, it is possible to notice inflections at $\sim$ 205 K in heating and cooling protocols (LPCMO's COO transition temperature, T$_{COO}$, see figures \hyperref[fig.Panel_3]{\ref*{fig.Panel_3} (a)} inset), and at $\sim$ 91 K and $\sim$ 63 K while warming and cooling, respectively. The latter corresponds to T$_C^W$ and T$_C^C$, respectively. These transition temperatures were determined from the maximum and minimum points of the first derivative of the moment with respect to temperature, as shown in figure \hyperref[fig.dMdT]{\ref*{fig.dMdT} (a)} in the Supplementary Materials \ref{Sup.Mat.Data}.

From the initial magnetization, denoted as M$_i$, and the final magnetization, denoted as M$_f$, at temperature T$_B$ preceding and following each thermal cycle, distinct regimes can be observed in figure \hyperref[fig.Panel_3]{\ref*{fig.Panel_3} (b)}. For T$_T$ $\leq$ 80 K (not shown), the final magnetization is approximately  equal to the initial one, with this difference decreasing as T$_T$ approaches 80 K. For T$_T$ $\geq$ 90 K, the final magnetization begins to be smaller than the initial one until both values become approximately equal at T$_T$ = 150 K. This persists until T$_T$ $\sim$ 190 K, from where the difference between M$_f$ and M$_i$ begins to increase again until T$_T$ $\geq$ 210 K, when a considerable jump in M$_f$ compared to M$_i$ is observed. As T$_T$ increases, the difference between the final and initial magnetization approaches zero again.

In panel \hyperref[fig.Panel_3]{\ref*{fig.Panel_3} (c) and (d)} resistivity curves as a function of temperature ($\rho$ - T) were collected by cooling and warming between 300 and 10 K, together with $\rho$ - T curves between T$_B$ and values of interest of T$_T$ are shown, analogously to figures \hyperref[fig.Panel_3]{\ref*{fig.Panel_3} (a) and (b)}. The resistivity measurements show an insulator-metal transition at approximately 120 K during cooling (T$_{IM}$). During heating, a metal-insulator transition occurs at around 153 K (T$_{MI}$). The positions of these were determined by the sign of the first derivative of the resistivity as a function of temperature and can be seen in figure \hyperref[fig.dMdT]{\ref*{fig.dMdT} (b)} in the Supplementary Materials \ref{Sup.Mat.Data}.

\begin{figure}[htbp!]
    \centering
    \includegraphics[width=.75\linewidth]{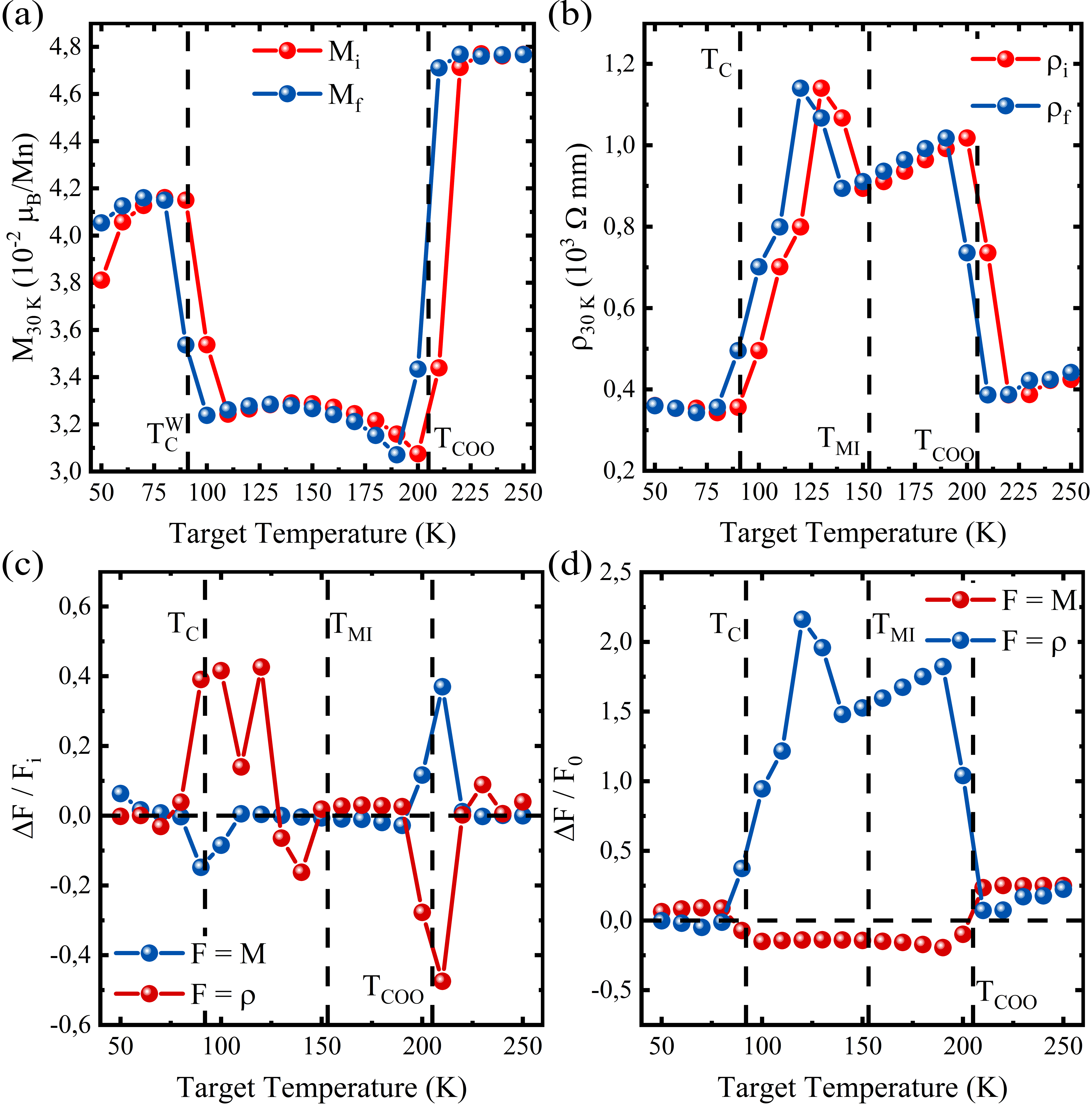}
    \caption{Figures (a) and (b) shows the magnetization and resistivity values at T$_B$ before (M$_i$ and $\rho_i$) and after (M$_f$ and $\rho_f$) each thermal cycle. Figures (c) and (d) shows the normalized difference, $(F_F-F)/F$, between final ($F_F$) and initial ($F$) magnetization (M) and resistivity ($\rho$) values for each individual cycle ($F=F_i$) and in respect to the initial values in T$_B$ = 30 K before thermal cycling commencing ($F=F_0$), respectively. The vertical dashed lines correspond to the respective magnetic and electrical transitions of the compound and the dotted line indicate T$_B$.}
    \label{fig.Panel_4}
\end{figure}

By comparing the thermal dependence of magnetic (figures \hyperref[fig.Panel_3]{\ref*{fig.Panel_3} (a) and (b)}) and  electrical  (\hyperref[fig.Panel_3]{\ref*{fig.Panel_3} (c) and (d)}) properties, it is possible to see that the final temperature at which the decrease (or increase) in final magnetization in relation to the initial magnetization after each cycle corresponds to the increase (or decrease) in final resistivity in relation to the initial resistivity. The figures \hyperref[fig.Panel_4]{\ref*{fig.Panel_4} (a) and (b)} illustrates this comparable and complementary behavior with M and $\rho$ measured at T$_B$ before (M$_i$ and $\rho_i$) and after (M$_f$ and $\rho_f$) each thermal cycle. The aforementioned correlation between M and $\rho$ is immediately observed. Furthermore, both cases show a dependence on the final state after each cycle with the transition temperatures T$_C$, T$_{MI}$ and T$_{COO}$, as in the intensity and width of the JT distortion modes.

Figures \hyperref[fig.Panel_4]{\ref*{fig.Panel_4} (c) and (d)} shows, analogously to figures \hyperref[fig.Panel_2]{\ref*{fig.Panel_2} (e) and (f)}, the normalized change ($F_F - F)/F$ in M and $\rho$ at T$_B$ with respect to the individual cycles (figure \hyperref[fig.Panel_4]{\ref*{fig.Panel_4} (c)}, $F=F_i$) and with respect to the initial values after cooling to T$_B$ and before starting the thermal cycles (figure \hyperref[fig.Panel_4]{\ref*{fig.Panel_4} (d)}, $F=F_0$). These figures, especially figures \hyperref[fig.Panel_4]{\ref*{fig.Panel_3} (a) and (b)}, show once again how T$_C$ and T$_{COO}$ are the main factors responsible for the behavior presented thus far. In figure \hyperref[fig.Panel_4]{\ref*{fig.Panel_4} (d)} it is possible to see a two-fold resistivity gain between T$_C$ and T$_{COO}$, when compared to the initial value. However, sample magnetization does not show quantitative agreement with these changes, producing only an approximate quarter in the same T$_T$ interval. This is possibly related to the percolative nature of the FMM regions, which causes the non compatible changes in resistivity and magnetization. 

\section*{Discussion}

Figures \hyperref[fig.Panel_2.2]{\ref*{fig.Panel_2.2} (e) and (f)} and figures \hyperref[fig.Panel_3]{\ref*{fig.Panel_3} (c) and (d)} show an clear agreement with each other. In the four cases (I$_{AS-JT}$, I$_{S-JT}$, M, and $\rho$), it is possible to see that the change in T$_B$ begins when the system is cycled near T$_C$. As T$_T$ moves away from T$_C$, the difference between the configurations before and after each cycle begins to diminish. Regarding the intensity of the JT modes and $\rho$, a distinct dependence on T$_{MI}$ is evident, which is not prominently observed in the behavior of magnetization. For T$_T$ $\sim$ T$_{MI}$, the difference between the initial and final states of the cycles tends towards zero, indicating the onset of the stable regime, as evidenced in previous figures, specifically \hyperref[fig.Panel_2.2]{\ref*{fig.Panel_2.2} (e) and (f)} and \hyperref[fig.Panel_4]{\ref*{fig.Panel_4} (c) and (d)}. The symmetry exhibited by the behavior of M and $\rho$ around T$_{COO}$, figures \hyperref[fig.Panel_4]{\ref*{fig.Panel_4} (a) and (b)}, underscores the validity of the prior interpretation of T$_{COO}$ as the system's reset temperature. 

Hence, Raman spectroscopy indicates both irreversible and reversible shifts in the FMM cluster distribution within the AFMI-COO matrix. An increase in resistivity correlates with reduced magnetization and heightened JT distortion modes, more pronounced in the insulating phase due to JT-polaronic transport and charge localization \cite{Billinge2002,Kim}. This suggests a reduction, stabilization, and eventual return of the clusters to initial values at T$_B$ as a function of T$_T$. More precisely, within the widely known percolation scenario for this compound \cite{Uehara}, thermal cycling changes the dimensionality of the percolation paths formed at T$_B$ according to the material's thermal history. Since the change in the resistivity of the system is represented by a maximum increase (at T$_C$ $<$ T$_T$ $<$ T$_{COO}$) of about three times, it cannot be inferred that the pre-formed percolation paths are \textit{completely} destroyed during the initial cooling of the sample to T$_B$, prior to the execution of thermal cycles. The formation of these, when cooling from 300 K, can be seen in figure \hyperref[fig.Panel_3]{\ref*{fig.Panel_3} (c)}, with the $\sim$ 10$^3$ resistivity drop after the T$_{IM}$ transition at around 120 K.

\begin{figure}[t!]
    \centering
    \includegraphics[width=\linewidth]{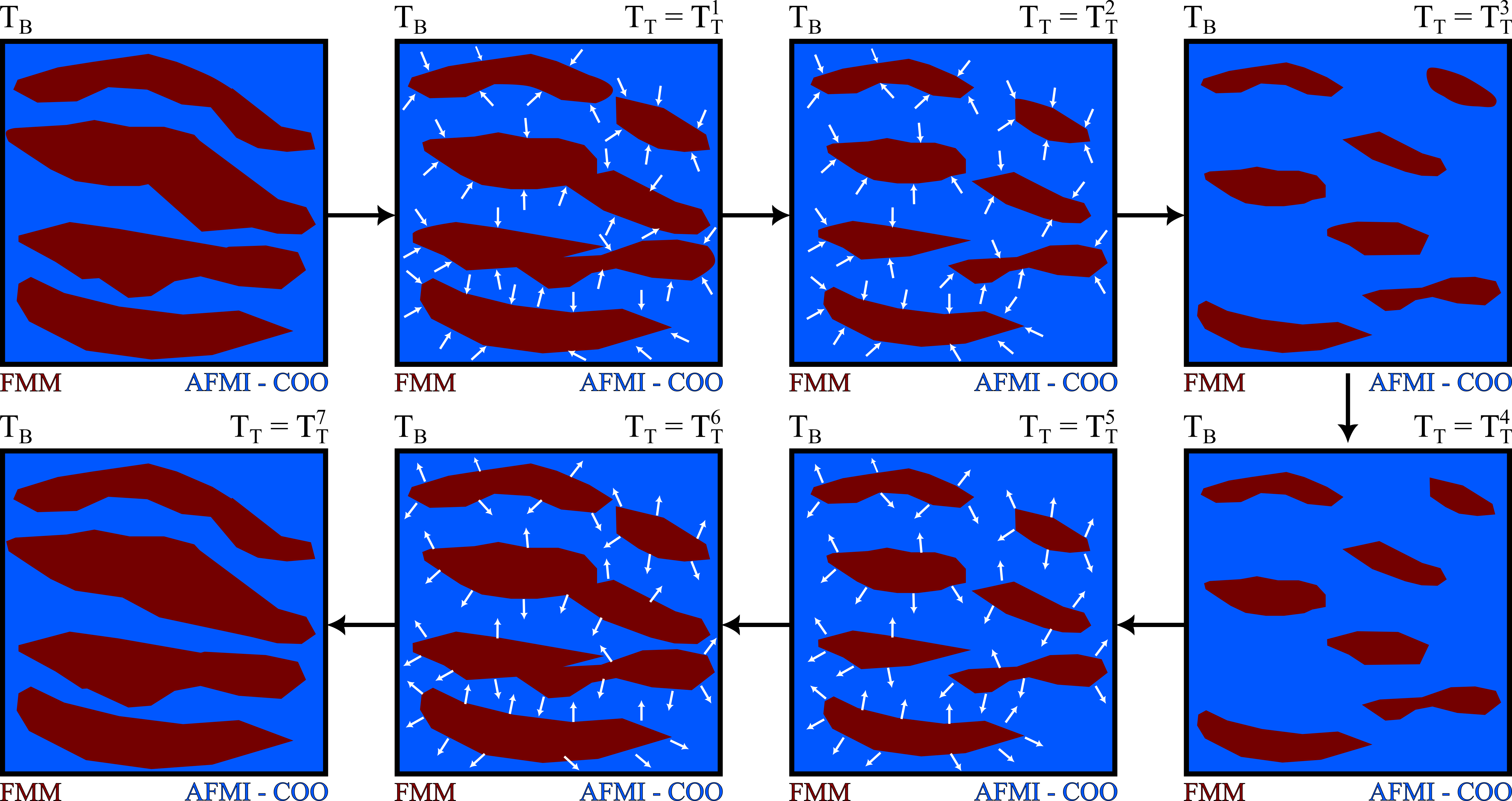}
    \caption{A schematic and qualitative representation of the changes in the FMM regions in the AFMI-COO matrix at T$_B$ after temperature cycling to T$_T$ is shown, illustrating the observed phenomena.}
    \label{fig.Ciclos.PS_Evo}
\end{figure}

In figure \hyperref[fig.Ciclos.PS_Evo]{\ref*{fig.Ciclos.PS_Evo}} a semi-quantitative and illustrative interpretation of the DPS reconfiguration scenario, as a function of the target temperature, proposed by all the presented data, is constructed. As the temperature is varied cyclically, the relative size of the FMM regions, percolated among themselves, begins to decrease at T$_B$ when T$_T$ exceeds T$_C$. In this process, it is also possible that a \textit{partial} disruption of some percolation paths occurs. For values of T$_T$ between T$_C$ and T$_{COO}$, the behavior stabilizes, so that the decrease in the FMM regions ceases. This is exemplified in figure \hyperref[fig.Ciclos.PS_Evo]{\ref*{fig.Ciclos.PS_Evo}} for T$_T$ = T$_T^3$ and T$_T^4$. For T$_T$ $\geq$ T$_{COO}$ the behavior is reversed, T$_T^6$, and the initial regime at T$_B$ is recovered, T$_T$ $\geq$ T$_T^6$.

Our results suggest the occurrence of a non-volatile resistive-switching (RS) in the sample as a consequence of cumulative thermal cycling around the T$_C$ $<$ T$_T$ $<$ T$_{MI}$ region. In contrast to conventional RS studies, where such effects are typically induced by current and/or voltage, the observed phenomenon here is entirely thermally driven, utilizing the material's inherent transitions for both irreversible and reversible modulation of resistivity. Moreover, it is demonstrated how the resistive switching (RS) phenomenon can be associated with the local structure of the sample. More specifically, this switching is found to involve alterations in the degree of structural order and the relative abundance of the antisymmetric (AS) and symmetric Jahn-Teller (S-JT) distorted regions. The ability to induce these effects through localized heating via a light source, while concurrently analyzing the resulting changes with Raman spectroscopy, holds substantial significance for the advancement of resistive devices and neuromorphic computing. This is attributed to the sample's capacity to store information structurally, magnetically, and electrically. Furthermore, it introduces an innovative method for controlling the phase separation (PS) regime and inducing resistive switching (RS) in complex oxides. Additional investigations parallel to those presented in this study, particularly focusing on manganites or other strongly correlated oxides known to exhibit resistive-switching (RS) responses, are highly desirable. This is especially true for studies where light serves as the primary mechanism of action.

\section*{Conclusions}

We have shown how thermally cycling a prototypical phase separated manganite between a base temperature deep in its dynamic PS regime and through its different magnetic, electric and structural transitions can provoke irreversible and reversible changes in such properties. Raman scattering has shown how such an effect is accompanied by the cyclic change in the sample's JT distortions and vibrational modes. The material has, after almost instantaneous cooling back to T$_B$, the capability of partially remembering the state to which it was heated. These local structural changes have been correlated to similar and complementary responses in the sample's magnetization and resistivity, resulting in a conjugated scenario of control and manipulation of its PS dynamics through its thermal history. 

In conclusion, the study demonstrates the potential of laser-induced thermal cycling to induce non-volatile structure/resistive-switching in complex oxides, leveraging the intrinsic phase transitions of the material. It was observed that the resistive-switching effect is thermally driven and correlates with changes in structural distortion order. This work introduces a novel method for controlling phase separation and resistive properties through localized laser heating and characterizing the previously laser-heated regions by Raman spectroscopy. Such findings have significant implications for the development of advanced resistive devices and neuromorphic systems, showcasing the material's ability to store and manage information structurally, magnetically, and electrically. Future investigations in similar oxide materials could further elucidate the role of localized heating and structural modulation in resistive-switching phenomena.

Acknowledgments -- This study was financed in part by the Coordenação de Aperfeiçoamento de Pessoal de Nível Superior – Brasil (CAPES) – Finance Code 001 and by the São Paulo Research Foundation (FAPESP), grant No. 2025/16064-5, 2024/00998-6 and 2023/17024-1. We also acknowledge the support of the INCT project Advanced Quantum Materials, involving CNPq (Proc. 408766/2024-7), FAPESP, and CAPES. This research used the facilities of the Brazilian Synchrotron Light Laboratory (LNLS), part of the Brazilian Center for Research in Energy and Materials (CNPEM), a private non-profit organization under the supervision of the Brazilian Ministry of Science, Technology, and Innovations (MCTI). The PAINEIRA beamline staff, especially Dra. Flávia Regina Estrada, are acknowledged for their assistance during the experiments 20250744. J.G. Ramirez and D. Carranza-Célis acknowledge support from Facultad de Ciencias and Vicerrectoría de investigaciones Universidad de los Andes, Convocatoria Facultad de Ciencias, Project No. INV-2023-178-2978. G. B. Gomide acknowledges the support of FAPESP under grant No. 2024/03819-5. M. Knobel and J.G. Ramirez acknowledge the support of FAPESP under grant No. 2022/16626-5.

\printbibliography

@article{Diego24,
  title={Low-temperature paramagnetic phase reentrance in praseodymium-doped manganites},
  author={Carranza-Celis, Diego and Wolowiec, Christian T and Basaran, Ali C and Salev, Pavel and Schuller, Ivan K and Ramirez, Juan Gabriel},
  journal={Physical Review Materials},
  volume={8},
  number={5},
  pages={054401},
  year={2024},
  publisher={APS}
}

@article{Iliev2003,
  title={Role of Jahn-Teller disorder in Raman scattering of mixed-valence manganites},
  author={Iliev, MN and Abrashev, MV and Popov, VN and Hadjiev, VG},
  journal={Physical Review B},
  volume={67},
  number={21},
  pages={212301},
  year={2003},
  publisher={APS}
}

@article{Carron,
  title={Raman phonons as a probe of disorder, fluctuations, and local structure in doped and undoped orthorhombic and rhombohedral manganites},
  author={Mart{\'\i}n-Carr{\'o}n, L and De Andres, A and Mart{\'\i}nez-Lope, MJ and Casais, MT and Alonso, JA},
  journal={Physical Review B},
  volume={66},
  number={17},
  pages={174303},
  year={2002},
  publisher={APS}
}

@article{Merten,
  title={Magnetic-field-induced suppression of Jahn-Teller phonon bands in (La$_{0.6}$Pr$_{0.4}$)$_{0.7}$Ca$_{0.3}$MnO$_3$: the mechanism of colossal magnetoresistance shown by Raman spectroscopy},
  author={Merten, Sebastian and Shapoval, O and Damaschke, B and Samwer, K and Moshnyaga, Vasily},
  journal={Scientific Reports},
  volume={9},
  number={1},
  pages={2387},
  year={2019},
  publisher={Nature Publishing Group UK London}
}

@article{Kim,
  title={Raman scattering studies of the temperature-and field-induced melting of charge order in La$_x$Pr$_y$Ca$_{1-x-y}$MnO$_3$},
  author={Kim, M and Barath, H and Cooper, SL and Abbamonte, P and Fradkin, E and R{\"u}bhausen, M and Zhang, CL and Cheong, S-W},
  journal={Physical Review B},
  volume={77},
  number={13},
  pages={134411},
  year={2008},
  publisher={APS}
}

@article{Uehara,
  title={Percolative phase separation underlies colossal magnetoresistance in mixed-valent manganites},
  author={Uehara, M and Mori, S and Chen, CH and Cheong, S-W},
  journal={Nature},
  volume={399},
  number={6736},
  pages={560--563},
  year={1999},
  publisher={Nature Publishing Group UK London}
}

@article{Sacanell2004,
  title={Low temperature irreversibility induced by thermal cycles on two prototypical phase separated manganites},
  author={Sacanell, J and Quintero, M and Curiale, J and Garbarino, G and Acha, C and Freitas, RS and Ghivelder, L and Polla, G and Leyva, G and Levy, P and others},
  journal={Journal of Alloys and Compounds},
  volume={369},
  number={1-2},
  pages={74--77},
  year={2004},
  publisher={Elsevier}
}

@article{Sacanell2018,
  title={Thermal cycling effects on static and dynamic properties of a phase separated manganite},
  author={Sacanell, Joaqu{\'\i}n and Sievers, Bernardo and Quintero, Mariano and Granja, Leticia and Ghivelder, Luis and Parisi, Francisco},
  journal={Journal of Magnetism and Magnetic Materials},
  volume={456},
  pages={212--216},
  year={2018},
  publisher={Elsevier}
}

@article{Levy2002,
  title={Nonvolatile magnetoresistive memory in phase separated La$_{0.325}$Pr$_{0.300}$Ca$_{0.375}$ MnO$_3$},
  author={Levy, Pablo and Parisi, Francisco and Quintero, Mariano and Granja, L and Curiale, J and Sacanell, J and Leyva, G and Polla, G and Freitas, RS and Ghivelder, L},
  journal={Physical Review B},
  volume={65},
  number={14},
  pages={140401},
  year={2002},
  publisher={APS}
}

@article{Quintero2007,
  title={Mechanism of electric-pulse-induced resistance switching in manganites},
  author={Quintero, M and Levy, P and Leyva, Adelma Graciela and Rozenberg, Marcelo Javier},
  journal={Physical Review Letters},
  volume={98},
  number={11},
  pages={116601},
  year={2007},
  publisher={APS}
}

@article{Ivan2024,
  title={Electrical Control of Magnetic Resonance in Phase Change Materials},
  author={Chen, Tian-Yue and Ren, Haowen and Ghazikhanian, Nareg and Hage, Ralph El and Sasaki, Dayne Y and Salev, Pavel and Takamura, Yayoi and Schuller, Ivan K and Kent, Andrew D},
  journal={Nano Letters},
  volume={24},
  number={37},
  pages={11476--11481},
  year={2024},
  publisher={ACS Publications}
}

@article{Granado1998,
  title={Phonon Raman scattering in R$_{1-x}$A$_x$MnO$_{3+\delta}$ (R= La, Pr; a= Ca, Sr)},
  author={Granado, E and Moreno, NO and Garcia, A and Sanjurjo, JA and Rettori, C and Torriani, I and Oseroff, SB and Neumeier, JJ and McClellan, KJ and Cheong, S-W and others},
  journal={Physical Review B},
  volume={58},
  number={17},
  pages={11435},
  year={1998},
  publisher={APS}
}

@article{Garcia2011,
  title={Multilevel hierarchy of phase separation processes in La$_{5/8-y}$Pr$_y$Ca$_{3/8}$MnO$_3$},
  author={Garc{\'\i}a-Mu{\~n}oz, JL and Collado, A and Aranda, MAG and Ritter, C},
  journal={Physical Review B},
  volume={84},
  number={2},
  pages={024425},
  year={2011},
  publisher={APS}
}

@article{Granado2000,
  title={Order-disorder in the Jahn-Teller transition of LaMnO$_3$: A Raman scattering study},
  author={Granado, Eduardo and Sanjurjo, Jose Antonio and Rettori, Carlos and Neumeier, JJ and Oseroff, SB},
  journal={Physical Review B},
  volume={62},
  number={17},
  pages={11304},
  year={2000},
  publisher={APS}
}

@article{Iliev1998,
  title={Raman spectroscopy of orthorhombic perovskitelike YMnO$_3$ and LaMnO$_3$},
  author={Iliev, MN and Abrashev, MV and Lee, H-G and Popov, VN and Sun, YY and Thomsen, Ch and Meng, RL and Chu, CW},
  journal={Physical Review B},
  volume={57},
  number={5},
  pages={2872},
  year={1998},
  publisher={APS}
}

@article{Liarokapis1999,
  title={Local lattice distortions and Raman spectra in the La$_{1-x}$Ca$_x$MnO$_3$ system},
  author={Liarokapis, E and Leventouri, Th and Lampakis, D and Palles, D and Neumeier, JJ and Goodwin, DH},
  journal={Physical Review B},
  volume={60},
  number={18},
  pages={12758},
  year={1999},
  publisher={APS}
}

@book{li2008synthesis,
  title={Synthesis of CMR manganites and ordering phenomena in complex transition metal oxides},
  author={Li, Haifeng},
  volume={4},
  year={2008},
  publisher={Forschungszentrum J{\"u}lich}
}

@article{Ahn2004,
  title={Strain-induced metal--insulator phase coexistence in perovskite manganites},
  author={Ahn, KH and Lookman, T and Bishop, AR},
  journal={Nature},
  volume={428},
  number={6981},
  pages={401--404},
  year={2004},
  publisher={Nature Publishing Group UK London}
}

@article{Podzorov2001,
  title={Martensitic accommodation strain and the metal-insulator transition in manganites},
  author={Podzorov, V and Kim, BG and Kiryukhin, V and Gershenson, ME and Cheong, SW},
  journal={Physical Review B},
  volume={64},
  number={14},
  pages={140406},
  year={2001},
  publisher={APS}
}

@article{Ghivelder2005,
  title={Dynamic phase separation in La$_{5/8-y}$Pr$_y$Ca$_{3/8}$MnO$_3$},
  author={Ghivelder, L and Parisi, F},
  journal={Physical Review B},
  volume={71},
  number={18},
  pages={184425},
  year={2005},
  publisher={APS}
}

@article{Dhakal2007,
  title={Effect of strain and electric field on the electronic soft matter in manganite thin films},
  author={Dhakal, Tara and Tosado, Jacob and Biswas, Amlan},
  journal={Physical Review B},
  volume={75},
  number={9},
  pages={092404},
  year={2007},
  publisher={APS}
}

@article{Radaelli1997,
  title={Charge, orbital, and magnetic ordering in La$_{0.5}$Ca$_{0.5}$MnO$_3$},
  author={Radaelli, PG and Cox, DE and Marezio, M and Cheong, SW},
  journal={Physical Review B},
  volume={55},
  number={5},
  pages={3015},
  year={1997},
  publisher={APS}
}

@article{Deng2024,
  title={Polarization-dependent photoinduced metal--insulator transitions in manganites},
  author={Deng, Lina and Zhang, Weiye and Lin, Hanxuan and Xiang, Lifen and Xu, Ying and Wang, Yadi and Li, Qiang and Zhu, Yinyan and Zhou, Xiaodong and Wang, Wenbin and others},
  journal={Science Bulletin},
  volume={69},
  number={2},
  pages={183--189},
  year={2024},
  publisher={Elsevier}
}

@article{Ogimoto2005,
  title={Strain-induced crossover of the metal-insulator transition in perovskite manganites},
  author={Ogimoto, Y and Nakamura, M and Takubo, N and Tamaru, H and Izumi, M and Miyano, K},
  journal={Physical Review B},
  volume={71},
  number={6},
  pages={060403},
  year={2005},
  publisher={APS}
}

@article{Collado2003,
  title={Room Temperature Structural and Microstructural Study for the Magneto-Conducting La$_{5/8-x}$Pr$_x$Ca$_{3/8}$MnO$_3$ (0$<$x$<$ 5/8) Series},
  author={Collado, Juan Antonio and Frontera, Carlos and Garc{\'\i}a-Mu{\~n}oz, Jos{\'e} Luis and Ritter, Clemens and Brunelli, M and Aranda, Miguel AG},
  journal={Chemistry of materials},
  volume={15},
  number={1},
  pages={167--174},
  year={2003},
  publisher={ACS Publications}
}

@article{Moshnyaga2014,
  title={Intrinsic antiferromagnetic coupling underlies colossal magnetoresistance effect: Role of correlated polarons},
  author={Moshnyaga, V and Belenchuk, Alexandr and Huehn, Sebastian and Kalkert, Christin and Jungbauer, Markus and Lebedev, Oleg I and Merten, Sebastian and Choi, K-Y and Lemmens, Peter and Damaschke, Bernd and others},
  journal={Physical Review B},
  volume={89},
  number={2},
  pages={024420},
  year={2014},
  publisher={APS}
}

@article{Granado2001,
  title={Effects of phase separation on the magnetization, x-ray diffraction, and Raman scattering of (La$_{1-y}$Nd$_y$)$_{1-x}$Ca$_x$MnO$_3$ (y= 0, 0.5, 1.0; x= 1/3)},
  author={Granado, E and Garc{\'\i}a, A and Sanjurjo, JA and Rettori, C and Torriani, I},
  journal={Physical Review B},
  volume={63},
  number={6},
  pages={064404},
  year={2001},
  publisher={APS}
}

@article{Dediu2000,
  title={Jahn-Teller dynamics in charge-ordered manganites from Raman spectroscopy},
  author={Dediu, V and Ferdeghini, C and Matacotta, FC and Nozar, P and Ruani, G},
  journal={Physical Review Letters},
  volume={84},
  number={19},
  pages={4489},
  year={2000},
  publisher={APS}
}

@article{Abrashev1999,
  title={Raman Study of the Variations of the Jahn-Teller Distortions through the Metal--Insulator Transition in Magnetoresistive La$_{0.7}$Ca$_{0. 3}$MnO$_3$ Thin Films},
  author={Abrashev, MV and Avanov, VG and Iliev, MN and Chakalov, RA and Chakalova, RI and Thomsen, C},
  journal={Physica Status Solidi (B)},
  volume={215},
  number={1},
  pages={631--636},
  year={1999},
  publisher={Wiley Online Library}
}

@article{Adams2000,
  title={Charge ordering and polaron formation in the magnetoresistive oxide La$_{0.7}$Ca$_{0.3}$MnO$_3$},
  author={Adams, CP and Lynn, JW and Mukovskii, YM and Arsenov, AA and Shulyatev, DA},
  journal={Physical Review Letters},
  volume={85},
  number={18},
  pages={3954},
  year={2000},
  publisher={APS}
}

@article{Schulman2024,
  title={Manganite Memristive Devices: Recent Progress and Emerging Opportunities},
  author={Schulman, Alejandro and Huhtinen, Hannu and Paturi, Petriina},
  journal={Journal of Physics D: Applied Physics},
  year={2024},
  publisher={IOP Publishing}
}

@article{Paasonen2024,
  title={Scalable and environmentally friendly production of perovskite manganite thin films for neuromorphic applications},
  author={Paasonen, Ville MM and Angervo, Ilari and Antola, Anni and Huhtinen, Hannu and Paturi, Petriina},
  journal={Thin Solid Films},
  volume={798},
  pages={140381},
  year={2024},
  publisher={Elsevier}
}

@article{Jaman2025,
  title={Electrically Induced Negative Differential Resistance States Mediated by Oxygen Octahedra Coupling in Manganites for Neuronal Dynamics},
  author={Jaman, Azminul and Fratino, Lorenzo and Ahmadi, Majid and Rocco, Rodolfo and Kooi, Bart J and Rozenberg, Marcelo and Banerjee, Tamalika},
  journal={Advanced Functional Materials},
  pages={2419840},
  year={2025},
  publisher={Wiley Online Library}
}

@article{Zhang2020,
  title={Understanding the metal-to-insulator transition in La$_{1- x}$Sr$_x$CoO$_{3-\delta}$ and its applications for neuromorphic computing},
  author={Zhang, Shenli and Galli, Giulia},
  journal={Npj Computational Materials},
  volume={6},
  number={1},
  pages={170},
  year={2020},
  publisher={Nature Publishing Group UK London}
}

@article{Lahteenlahti2021,
  title={Electron Doping Effect in the Resistive Switching Properties of Al/Gd$_{1-x}$Ca$_x$MnO$_3$/Au Memristor Devices},
  author={Lahteenlahti, Ville and Schulman, Alejandro and Beiranvand, Azar and Huhtinen, Hannu and Paturi, Petriina},
  journal={ACS Applied Materials \& Interfaces},
  volume={13},
  number={15},
  pages={18365--18371},
  year={2021},
  publisher={ACS Publications}
}

@article{Laukhin1997,
  title={Pressure effects on the metal-insulator transition in magnetoresistive manganese perovskites},
  author={Laukhin, V and Fontcuberta, J and Garcia-Munoz, JL and Obradors, X},
  journal={Physical Review B},
  volume={56},
  number={16},
  pages={R10009},
  year={1997},
  publisher={APS}
}

@article{Baldini2012,
  title={Pressure induced magnetic phase separation in La$_{0.75}$Ca$_{0.25}$MnO$_3$ manganite},
  author={Baldini, Monica and Capogna, L and Capone, Massimo and Arcangeletti, Emanuele and Petrillo, Caterina and Goncharenko, I and Postorino, Paolo},
  journal={Journal of Physics: Condensed Matter},
  volume={24},
  number={4},
  pages={045601},
  year={2012},
  publisher={IOP Publishing}
}

@article{Lin2018,
  title={Unexpected intermediate state photoinduced in the metal-insulator transition of submicrometer phase-separated manganites},
  author={Lin, Hanxuan and Liu, Hao and Lin, Lingfang and Dong, Shuai and Chen, Hongyan and Bai, Yu and Miao, Tian and Yu, Yang and Yu, Weichao and Tang, Jing and others},
  journal={Physical Review Letters},
  volume={120},
  number={26},
  pages={267202},
  year={2018},
  publisher={APS}
}

@article{Sarma2004,
  title={Direct observation of large electronic domains with memory effect in doped manganites},
  author={Sarma, DD and Topwal, Dinesh and Manju, U and Krishnakumar, SR and Bertolo, M and La Rosa, S and Cautero, Giuseppe and Koo, TY and Sharma, PA and Cheong, S-W and others},
  journal={Physical Review Letters},
  volume={93},
  number={9},
  pages={097202},
  year={2004},
  publisher={APS}
}

@article{Ward2011,
  title={Dynamics of a first-order electronic phase transition in manganites},
  author={Ward, Thomas Z and Gai, Zheng and Guo, HW and Yin, LF and Shen, Jian},
  journal={Physical Review B},
  volume={83},
  number={12},
  pages={125125},
  year={2011},
  publisher={APS}
}

@article{Merten2019,
  title={Jahn-Teller reconstructed surface of the doped manganites shown by means of surface-enhanced Raman spectroscopy},
  author={Merten, S and Bruchmann-Bamberg, V and Damaschke, B and Samwer, K and Moshnyaga, Vasily},
  journal={Physical Review Materials},
  volume={3},
  number={6},
  pages={060401},
  year={2019},
  publisher={APS}
}

@article{Amelitchev2001,
  title={Structural and chemical analysis of colossal magnetoresistance manganites by Raman spectrometry},
  author={Amelitchev, VA and G{\"u}ttler, B and Gorbenko, O Yu and Kaul, AR and Bosak, AA and Ganin, A Yu},
  journal={Physical Review B},
  volume={63},
  number={10},
  pages={104430},
  year={2001},
  publisher={APS}
}

@article{GSAS,
  title={GSAS-II: the genesis of a modern open-source all purpose crystallography software package},
  author={Toby, Brian H and Von Dreele, Robert B},
  journal={Applied Crystallography},
  volume={46},
  number={2},
  pages={544--549},
  year={2013},
  publisher={International Union of Crystallography}
}

@inproceedings{PAINEIRA,
  title={PAINEIRA beamline at Sirius: an automated facility for polycrystalline XRD characterization},
  author={Estrada, FR and Barrett, DH and Ferreira, AI and Mauricio, JC and Rigamonti Jr, H and Meyer, BC and Tolentino, HCN and Westfahl Jr, H and Rodella, CB},
  booktitle={Journal of Physics: Conference Series},
  volume={2380},
  number={1},
  pages={012033},
  year={2022},
  organization={IOP Publishing}
}

@article{Gomide2025,
    author = {Gomide, Gabriel B. and Carranza-Celis, Diego and Kuhl, Guilherme and Knobel, Marcelo and Ramírez, Juan Gabriel and Muraca, Diego},
    title = {Voltage-tunable spin resonance in quantum phase-separated material},
    journal = {Applied Physics Letters Materials},
    volume = {13},
    number = {4},
    pages = {041122},
    year = {2025},
    abstract = {The voltage control of spin and charge degrees of freedom in complex materials is a cornerstone for the realization of advanced electronic devices with enhanced functionalities. Here, we demonstrate in situ indirect current control via the Joule effect of the spin resonance parameters in a phase-separated La5/8−yPryCa3/8MnO3 sample while simultaneously inducing resistive switching. By employing electron paramagnetic resonance (EPR) spectroscopy under an applied bias voltage, we observe sharp, reversible modifications in the EPR spectra—linewidth, resonance field, and intensity—concurrent with voltage-driven transitions between the ferromagnetic metallic (FMM) and paramagnetic charge-ordered (PM-CO) states. This real-time probing of spin resonance during resistive switching provides crucial insights into the interplay between spin, charge, and lattice degrees of freedom, elucidating the distinct roles of the FMM and PM-CO phases in the observed behavior. These findings pave the way for the development of novel spintronic and neuromorphic devices with voltage-tunable functionalities.},
    %issn = {2166-532X},
    %doi = {10.1063/5.0256253},
    %url = {https://doi.org/10.1063/5.0256253},
    %eprint = {https://pubs.aip.org/aip/apm/article-pdf/doi/10.1063/5.0256253/20498107/041122\_1\_5.0256253.pdf},
}

@article{Goff2004,
  title = {Charge ordering in half-doped manganites},
  author = {Goff, R. J. and Attfield, J. P.},
  journal = {Physical Review B},
  volume = {70},
  issue = {14},
  pages = {140404},
  numpages = {4},
  year = {2004},
  publisher = {American Physical Society},
  %doi = {10.1103/PhysRevB.70.140404},
  %url = {https://link.aps.org/doi/10.1103/PhysRevB.70.140404}
}

@article{Kiryukhin2000,
  title={Multiphase segregation and metal-insulator transition in single crystal La$_{5/8-y}$ Pr$_y$Ca$_{3/8}$MnO$_3$},
  author={Kiryukhin, V and Kim, BG and Podzorov, V and Cheong, S-W and Koo, TY and Hill, JP and Moon, I and Jeong, YH},
  journal={Physical Review B},
  volume={63},
  number={2},
  pages={024420},
  year={2000},
  publisher={APS}
}

@article{Lee2002,
  title={Optical evidence of multiphase coexistence in single crystalline (La, Pr, Ca)MnO$_3$},
  author={Lee, HJ and Kim, KH and Kim, MW and Noh, TW and Kim, BG and Koo, TY and Cheong, S-W and Wang, YJ and Wei, X},
  journal={Physical Review B},
  volume={65},
  number={11},
  pages={115118},
  year={2002},
  publisher={APS}
}

@incollection{Billinge2002,
  title={Polarons in Manganites; Now You See Them Now You Don’t},
  author={Billinge, Simon JL},
  booktitle={Physics of Manganites},
  pages={201--210},
  year={2002},
  publisher={Springer}
}

@article{li2022,
  title={Electronically phase separated nano-network in antiferromagnetic insulating LaMnO3/PrMnO3/CaMnO3 tricolor superlattice},
  author={Li, Qiang and Miao, Tian and Zhang, Huimin and Lin, Weiyan and He, Wenhao and Zhong, Yang and Xiang, Lifen and Deng, Lina and Ye, Biying and Shi, Qian and others},
  journal={Nature Communications},
  volume={13},
  number={1},
  pages={6593},
  year={2022},
  publisher={Nature Publishing Group UK London}
}

@article{Zhou2015,
  title={Evolution and control of the phase competition morphology in a manganite film},
  author={Zhou, Haibiao and Wang, Lingfei and Hou, Yubin and Huang, Zhen and Lu, Qingyou and Wu, Wenbin},
  journal={Nature Communications},
  volume={6},
  number={1},
  pages={8980},
  year={2015},
  publisher={Nature Publishing Group UK London}
}

@article{kundhikanjana2015,
  title={Direct imaging of dynamic glassy behavior in a strained manganite film},
  author={Kundhikanjana, Worasom and Sheng, Zhigao and Yang, Yongliang and Lai, Keji and Ma, Eric Yue and Cui, Yong-Tao and Kelly, Michael A and Nakamura, Masao and Kawasaki, Masashi and Tokura, Yoshinori and others},
  journal={Physical Review Letters},
  volume={115},
  number={26},
  pages={265701},
  year={2015},
  publisher={APS}
}

@article{Ward2009,
  title={Elastically driven anisotropic percolation in electronic phase-separated manganites},
  author={Ward, Thomas Z and Budai, John D and Gai, Zheng and Tischler, Jonathan Zachary and Yin, Lifeng and Shen, Jian},
  journal={Nature Physics},
  volume={5},
  number={12},
  pages={885--888},
  year={2009},
  publisher={Nature Publishing Group UK London}
}

@article{Kim2021,
  title={Raman imaging of ferroelastically configurable Jahn--Teller domains in LaMnO3},
  author={Kim, Yong-Jin and Park, Heung-Sik and Yang, Chan-Ho},
  journal={Npj Quantum Materials},
  volume={6},
  number={1},
  pages={62},
  year={2021},
  publisher={Nature Publishing Group UK London}
}

@article{Milloch2024,
  title={Mott resistive switching initiated by topological defects},
  author={Milloch, Alessandra and Figueruelo-Campanero, Ignacio and Hsu, Wei-Fan and Mor, Selene and Mellaerts, Simon and Maccherozzi, Francesco and Veiga, Larissa SI and Dhesi, Sarnjeet S and Spera, Mauro and Seo, Jin Won and others},
  journal={Nature Communications},
  volume={15},
  number={1},
  pages={9414},
  year={2024},
  publisher={Nature Publishing Group UK London}
}

@article{Kalcheim2020,
  title={Non-thermal resistive switching in Mott insulator nanowires},
  author={Kalcheim, Yoav and Camjayi, Alberto and Del Valle, Javier and Salev, Pavel and Rozenberg, Marcelo and Schuller, Ivan K},
  journal={Nature communications},
  volume={11},
  number={1},
  pages={2985},
  year={2020},
  publisher={Nature Publishing Group UK London}
}

\newpage

\beginsupplement
\section{Supplementary Materials.}\label{Supplementary Materials}

\section{Local temperature as a function of laser power}\label{Calibration}

\begin{figure}[H]
    \centering
    \includegraphics[width=\linewidth]{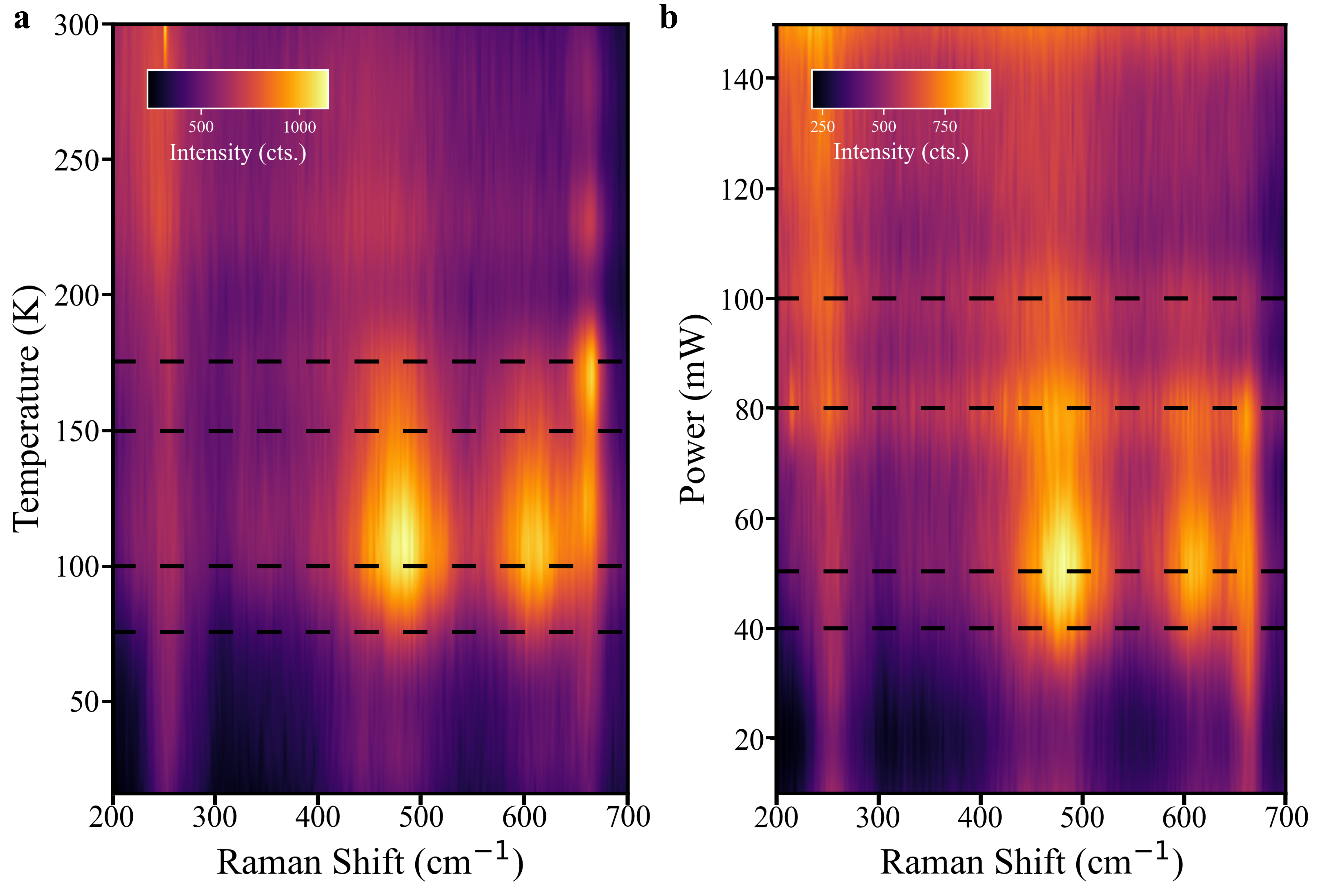}
    \caption{Intensity maps constructed by interpolating the intensity of the collected spectra as a function of temperature, \textbf{a}, and as a function of continuously increasing power, \textbf{b}. The horizontal dashed lines indicate the temperatures/powers at which inflections in the intensity of the structures occur at $\sim$ 480 $cm^{-1}$ and $\sim$ 614 $cm^{-1}$. The thermal Bose-Einstein factor was not excluded from any of the spectra.}
    \label{fig.Heat.Maps.P.T}
\end{figure}

Raman spectra as a function of temperature and laser power were collected for the sample $y$ = 0.35. The local sample temperature as a function of laser power was determined by directly comparing spectra at a fixed power (10 mW) with variable temperatures and at a fixed temperature (16 K) with variable powers. These will be referred to as spectra as  function of temperature and power, respectively. This comparison was based primarily on three factors: the similarity between the low-frequency regions, which is directly related to the thermal Bose-Einstein factor (and therefore to temperature); and the width and relative intensity of the three main spectral structures, centered at $\sim$ 250 $cm^{-1}$, $\sim$ 480 $cm^{-1}$, and $\sim$ 614 $cm^{-1}$. After an initial estimation, the fit parameters of the three structures from both sets of data were compared to verify their agreement. It is assumed that the increase in local temperature due to the increase in power should be linear or approximately linear.

\begin{figure}[H]
    \centering
    \includegraphics[width=\linewidth]{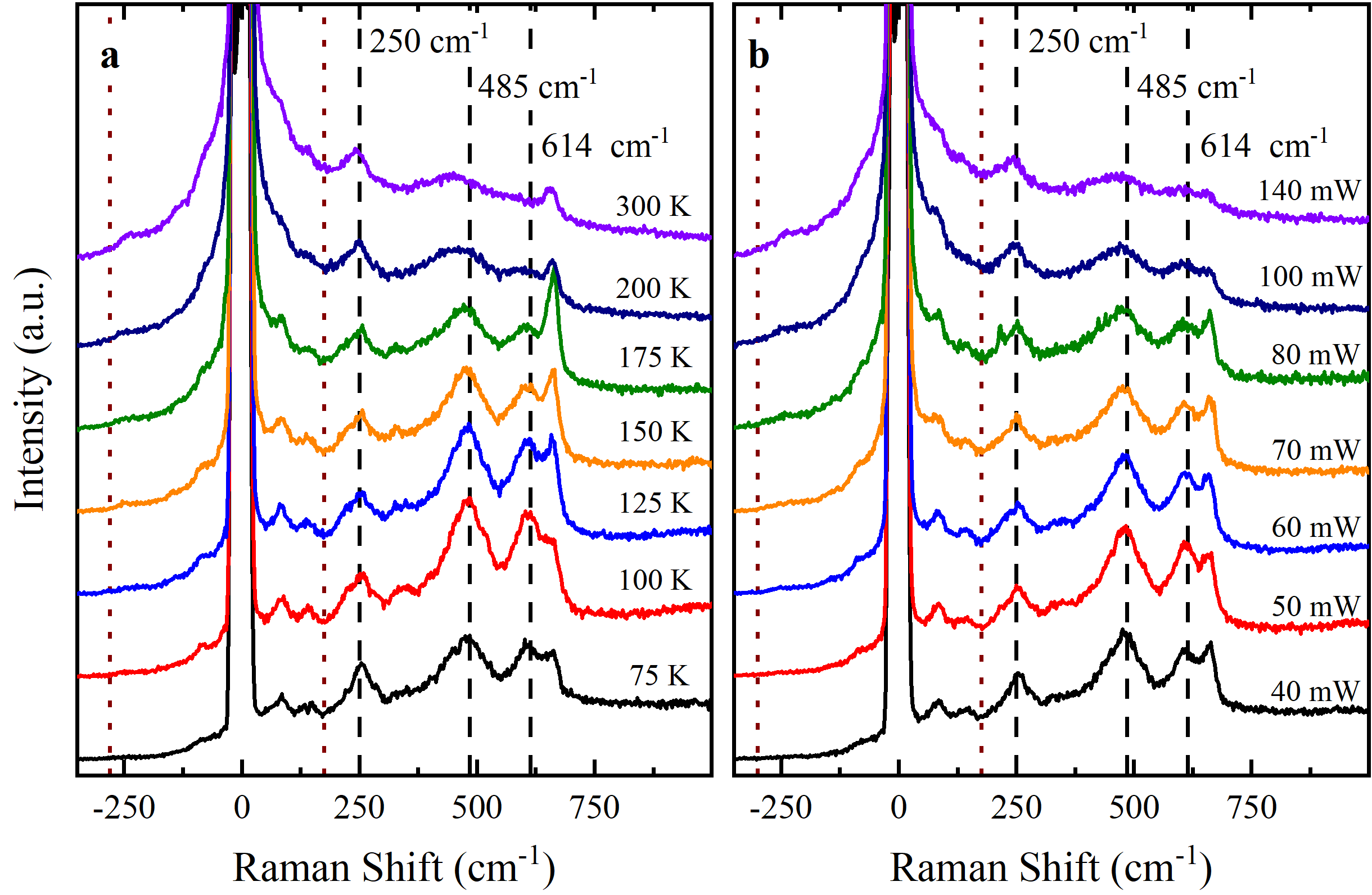}
    \caption{Raman spectra of the sample $y$ = 0.35 collected by heating with the temperature controller's heater, \textbf{a}, and by heating with the laser power, \textbf{b}. The spectra were vertically translated to align the lower limit of their anti-Stokes signals. The vertical dashed lines indicate the positions of the modes at 250, 485, and 614 cm$^{-1}$ and the vertical dotted lines delimit the low-frequency region, where the influence of the thermal Bose-Einstein factor is greatest.}
    \label{fig.PxT_Trans}
\end{figure}

Figures \hyperref[fig.Heat.Maps.P.T]{\ref*{fig.Heat.Maps.P.T} (a) and (b)} present intensity maps corresponding to the spectra as functions of temperature and power, respectively. Within these maps, the principal inflection points in the intensities of the structures at $\sim$ 480 $cm^{-1}$ and $\sim$ 614 $cm^{-1}$ are delineated by dashed horizontal lines. In Figure \hyperref[fig.Heat.Maps.P.T]{\ref*{fig.Heat.Maps.P.T} (a)}, it is apparent that there is an increase in the intensity of these structures at approximately 75 K. The maximum intensity is reached around 100 K. Then, after 125 K, the intensities of both structures begin to decrease. At 175 K there is an almost complete loss of intensity by the structure $\sim$ 614 $cm^{-1}$ and a large broadening, together with the loss of intensity, of the structure at $\sim$ 480 $cm^{-1}$. In Figure \hyperref[fig.Heat.Maps.P.T]{\ref*{fig.Heat.Maps.P.T} (b)}, it is possible to note that similar intensity inflections in the same structures of interest are observed at 40, 50, 80, and 100 mW, respectively. Therefore, it is possible to observe that there is a correspondence between the intensities of the mentioned regions for both heating protocols. It is important to note that for the data as a function of power, the intensity variation was accompanied by an increase in the background signal, which was not the same for all measurements. This, as well as possible increases in the intensity of elastic scattering, was also considered.

In figures \hyperref[fig.PxT_Trans]{\ref{fig.PxT_Trans} (a) and (b)}, some of the Raman spectra, vertically shifted, collected as a function of the heating produced by the cold finger heater and the increase in laser power, respectively, are shown. The spectra aligned horizontally in the two figures are those that present the greatest similarity to each other, considering the factors mentioned previously. The structures around 250, 485, and 614 cm$^{-1}$ are identified in the figures by vertical dashed lines, while the low-frequency region ($\sim$ - 300 cm$^{-1}$ to $\sim$ 200 cm$^{-1}$) is delimited by vertical dotted lines.

\begin{table}[ht]
\center
\begin{tabular}{ccc|cc}
\toprule
\makecell{Power (mW)}& \makecell{Cryo. \\ Temp. (K)} &  \makecell{Temp. \\ Fit. (K)} & \makecell{Adj. Cryo \\ Temp. (K)} &  \makecell{Adj. Cryo  \\ Temp. Fit (K)} \\
\midrule
10 & 16 & 20.9 (2) & 37 & 38 (4)\\
20 & - & 41.5 (4) & - & 59 (4)\\
30 & - & 62.8 (6) & - & 81 (4)\\
40 & 75 & 83.7 (8) & 96 & 102 (4)\\
50 & 100 & 105 (9) & 121 & 124 (5)\\
60 & 125 & 126 (1) & 146 & 145 (5)\\
70 & 150 & 146 (1) & 171 & 167 (5)\\
80 &  175 & 167 (2) & 196 & 188 (5)\\
90 & - & 188 (2) & - & 209 (6)\\
100 & 200  & 209 (2) & 221 & 231 (6)\\
110 & 225 & 230 (2) & 246 & 252 (6)\\
120 & 250 & 251 (2) & 271 & 274 (7)\\
130 & 275  & 272 (3) & 296 & 295 (7)\\
140 & 300 & 293 (3) & 321& 317 (7)\\
150 & - & 314 (3) & - &  338 (8)\\
\bottomrule
\end{tabular}
\caption[Temperature estimate as a function of laser power.]{Temperature estimate based on visual inspection and comparison of spectra, first two columns, and temperature obtained by linear fitting of the calibration, third column. The last two columns correspond to the adjusted cold finger temperature, assuming that 10 mW generates an average heating of 21.8 K. The numbers in parentheses correspond to the uncertainty of the last decimal place shown, obtained from the linear fit.}
\label{tab.1}
\end{table}

Hence, from the figure, it is feasible to establish the following equivalence between temperature and power: 40 mW $\sim$ corresponds to 75 K; 50 mW $\sim$ to 100 K; 60 mW $\sim$ to 125 K; 70 mW $\sim$ to 150 K; 80 mW $\sim$ to 175 K; 100 mW $\sim$ to 200 K; and 140 mW $\sim$ to 300 K. However, the groups of spectra do not align perfectly, introducing some degree of uncertainty in this determination, particularly as not all spectra exhibit identical backgrounds. These uncertainties are accounted for in the linear fit.

For temperatures and powers lower than 75 K and 40 mW, respectively, no differences(except for small variations in the background and in the intensity of elastic scattering) are observed. Consequently, it can be inferred that, within the power range of 10 mW $<$ to 40 mW $<$, the heating effect spans from 16 K to 50 K. Spectroscopic analyzes conducted at 16 K, 30 K, and 50 K corroborate the absence of distinguishable differences in the obtained spectra at these temperatures.

\begin{figure}[H]
    \centering
    \includegraphics[width=\linewidth]{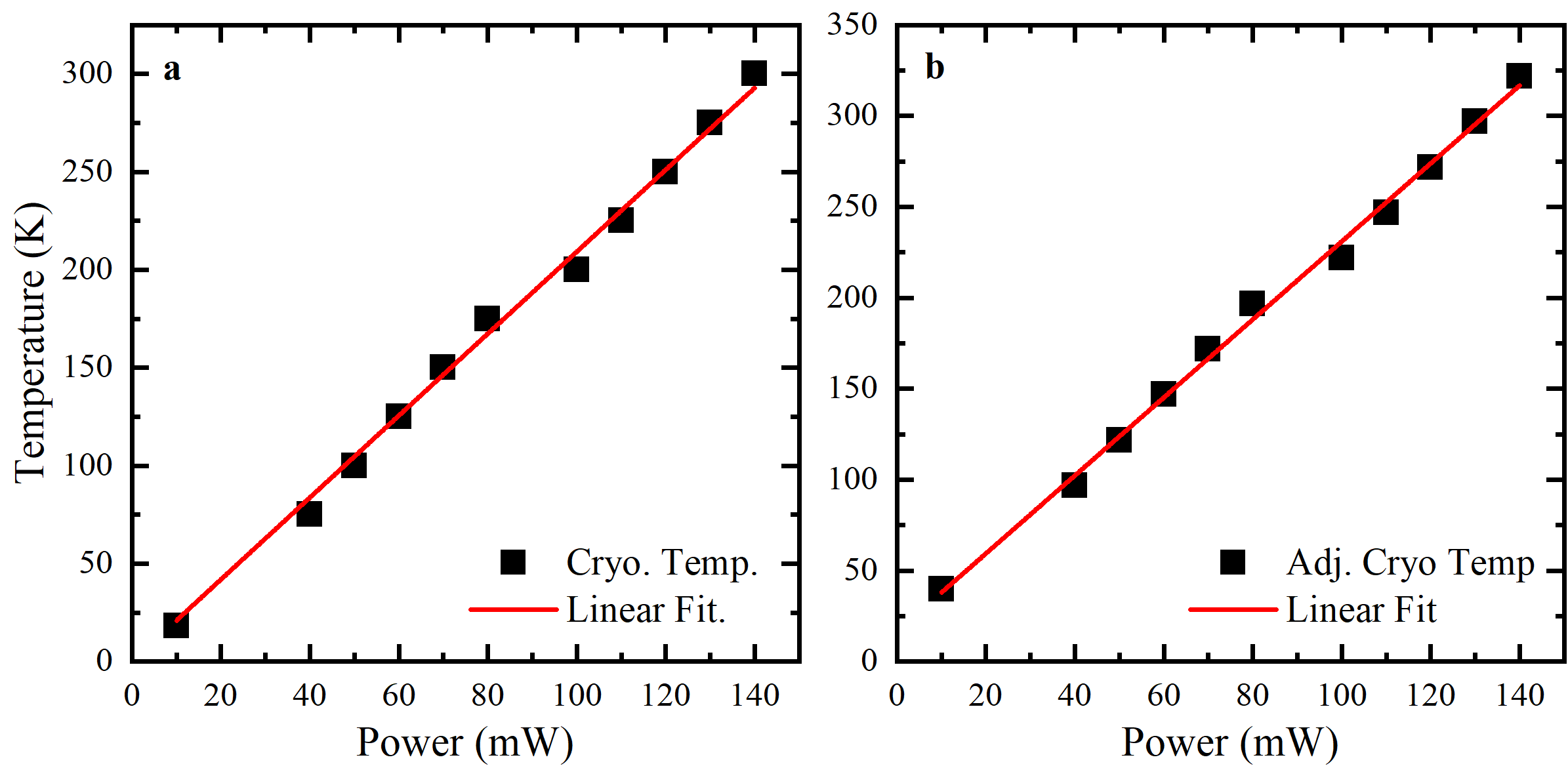}
    \caption[Linear fit of temperature as a function of power.]{Linear fit of temperature as a function of power for values determined by visual comparison between spectra. \textbf{a} fit of the second column in table \ref{tab.1} as a function of power and \textbf{b} fit of the fourth column in table \ref{tab.1} as a function of power.}
    \label{fig.linear.fit}
\end{figure}

The complete power-temperature designation is organized in table \ref{tab.1}. First, based on the first two columns of table \ref{tab.1},an approximately linear increase in temperature is observed with increasing laser power, and the data can be appropriately fitted to a linear equation. The best fit was obtained with the equation T(P) = 2.09 (0.02) P \footnote{If both fitting parameters were varied, the intercept becomes negative, more precisely $\sim$ -3 (5) K. Since this does not make physical sense (zero power implies negative temperatures), it was fixed to zero.}, as shown in figure \hyperref[fig.linear.fit]{\ref*{fig.linear.fit} (a)}. Substituting the values of power, P, into this equation, one obtains the third column of table \ref{tab.1}.

At first glance, one might assume that 10 mW does not generate any heating at the analyzed point.Nevertheless, it can be observed that under the approximation of a linear relationship, a power of $\Delta$P = 110 mW corresponds to a temperature of $\Delta$T = 232 K. Thus, $\Delta$T / $\Delta$P = 2.18 K/mW, which implies that 10 mW is equivalent to a heating of 21.8 K. Therefore, it is necessary to add this value to the temperatures provided by the temperature controller. Thus, the fourth column of table \ref{tab.1} is obtained. Fitting these points to a linear equation, the best fit is T(P) = 2.14 (0.04)P + 16 (4), shown in figure \hyperref[fig.linear.fit]{\ref*{fig.linear.fit} (b)}. The values obtained from this last fit can be found in the last column of table \ref{tab.1}. The errors in the third and fifth columns are based on the propagation of standard errors. The temperature values according to the linear fit, fifth column of table \ref{tab.1}, will be adopted for the data as a function of power.

\begin{figure}[H]
    \centering
    \includegraphics[width=\linewidth]{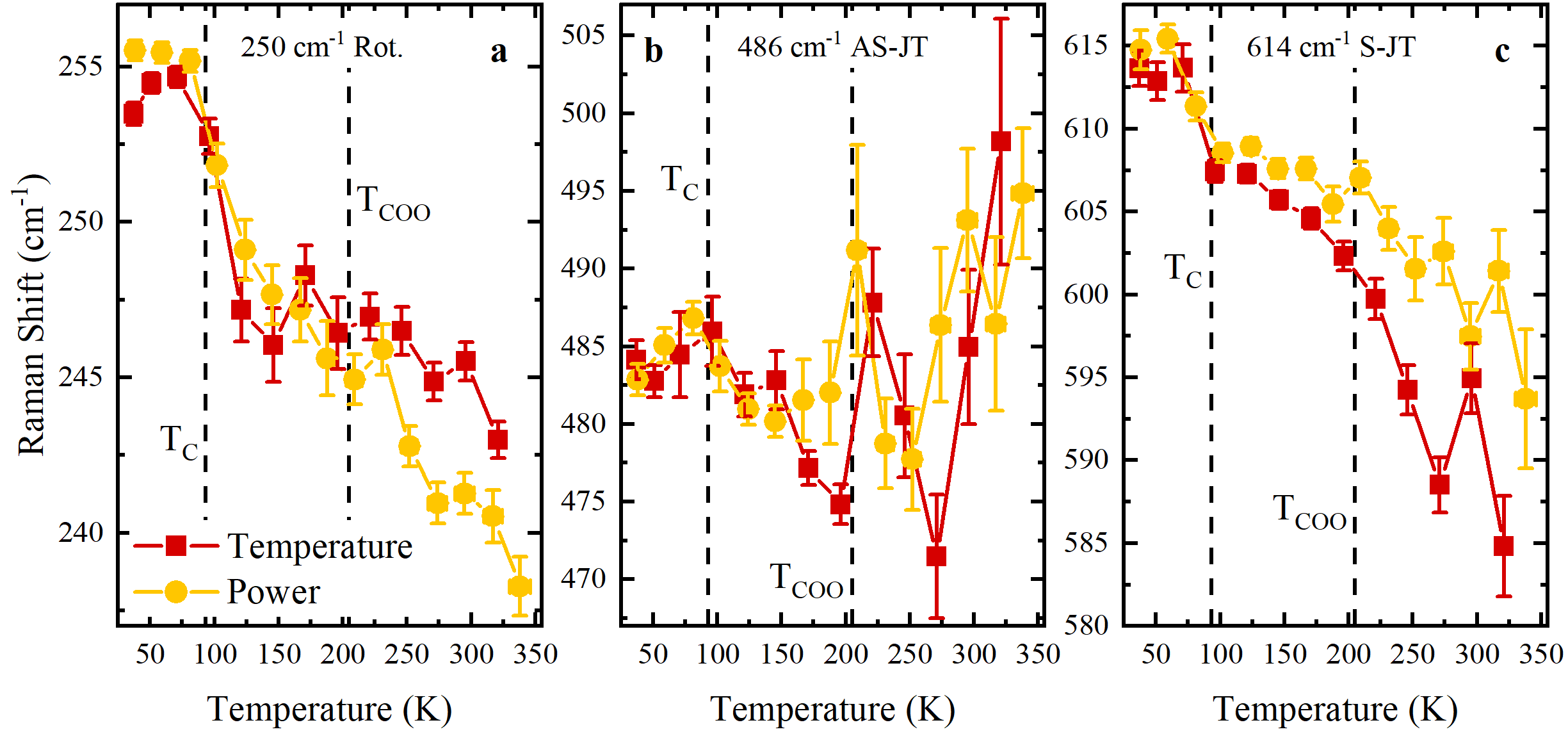}
    \caption{Vibrational mode centers centered at \textbf{a} 250 cm$^{-1}$, \textbf{b} 486 cm$^{-1}$ and \textbf{c} 614 cm$^{-1}$ collected as a function of temperature (red) and power (orange), where the temperature calibration performed is considered. The vertical dashed lines indicate the transition temperatures T$_{FPS}$ and T$_{OCO}$.}
    \label{fig.PxT.Centers}
\end{figure}

In figures \hyperref[fig.PxT.Centers]{\ref*{fig.PxT.Centers} (a), (b) and (c)} the centers of the modes at $\sim$ 250 cm$^{-1}$, $\sim$ 486 cm$^{-1}$ and $\sim$ 614 cm$^{-1}$ are shown, corresponding to rotation (Rot.), anti-symmetric Jahn-Teller distortion (AS-JT) and symmetric Jahn-Teller distortion (S-JT) modes, respectively, for the spectra collected as a function of temperature and power. Good agreement in the behavior and energy values for the three peaks can be observed over the entire temperature range and within the fitting error. Furthermore, the centers of the three structures show similar dependencies on the sample transition temperatures, T$_{FPS}$, at $\sim$ 93 K, and T$_{OCO}$, at $\sim$ 205 K.

\begin{figure}[!t]
    \centering
    \includegraphics[width=\linewidth]{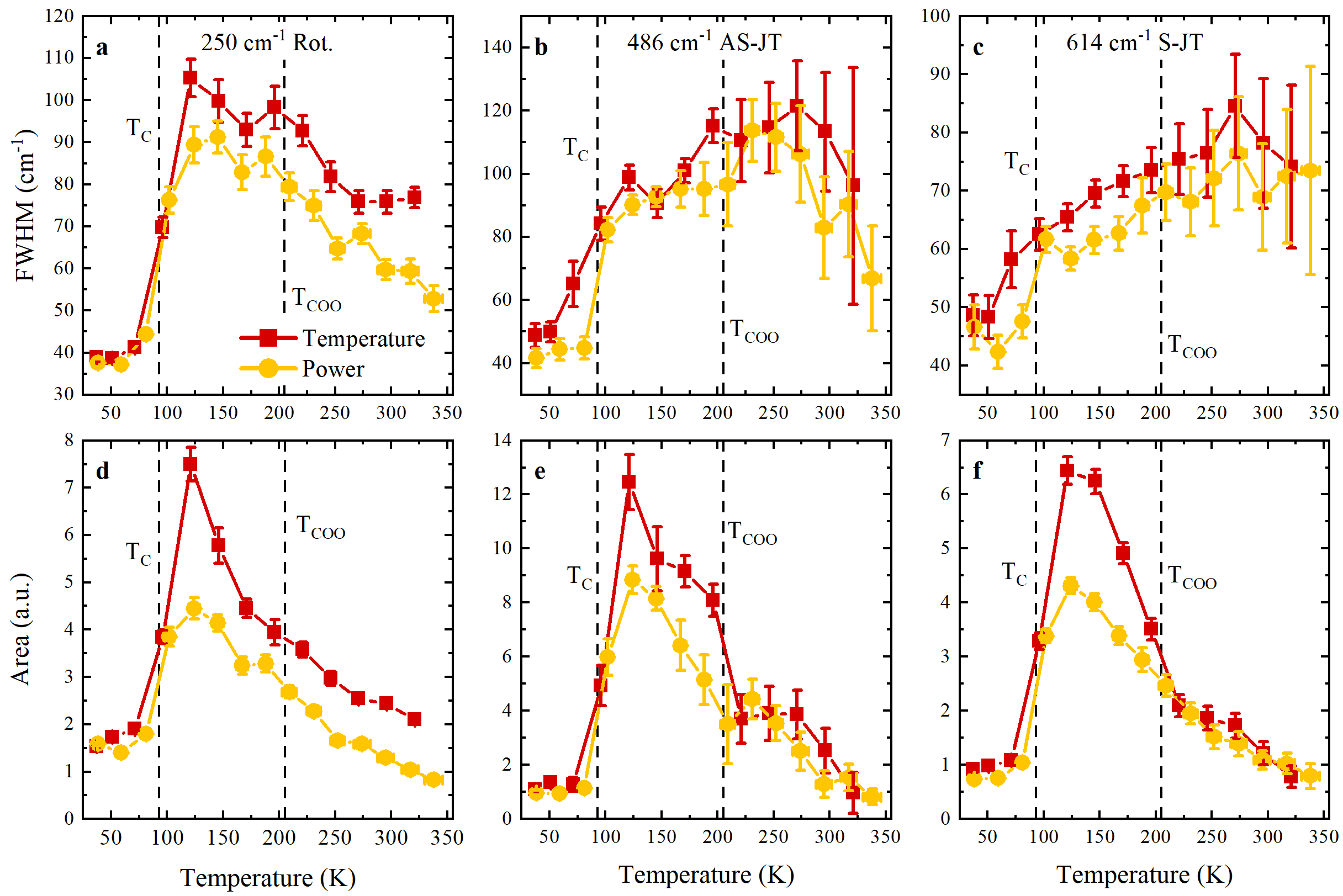}
    \caption[Width and area of vibrational modes of interest obtained from fitting collected Raman spectra as a function of temperature and power.]{Spectra fitting parameters as a function of temperature (red) and power (orange), where the temperature calibration performed is considered. In \textbf{a}, \textbf{b} and \textbf{c}, the FWHM of the modes at 250 cm$^{-1}$, 480 cm$^{-1}$ and 614 cm$^{-1}$ are shown, respectively, and in \textbf{d}, \textbf{e} and \textbf{f} the areas of the same structures. The vertical dashed lines indicate the transition temperatures T$_{FPS}$ and T$_{OCO}$.}
    \label{fig.PxT.FWHM_Area}
\end{figure}

In figures \hyperref[fig.PxT.FWHM_Area]{\ref*{fig.PxT.FWHM_Area} (a), (b) and (c)}, the FWHM of the modes at $\sim$ 250 cm$^{-1}$, $\sim$ 486 cm$^{-1}$, and $\sim$ 614 cm$^{-1}$, respectively, is shown. It can be observed that the width behavior of both heating protocols presents good agreement, as well as their dependence on the sample transition temperatures. In figures \hyperref[fig.PxT.FWHM_Area]{\ref*{fig.PxT.FWHM_Area} (d), (e) and (f)} the areas of the same modes are shown, respectively. Again, it is possible to observe a great agreement in both behavior and values between both two procedures for the three structures.

The differences observed for T $>$ T$_{OCO}$ in the energy of the rotational mode at $\sim$ 250 cm$^{-1}$ and for the S-JT distortion mode at $\sim$ 614 cm$^{-1}$, as shown in figures \hyperref[fig.PxT.Centers]{\ref*{fig.PxT.Centers} (c) and (f)}, and the discrepancies observed in the FWHM of the mode at $\sim$ 250 cm$^{-1}$ for T $>$ T$_{OCO}$ and of the mode at $\sim$ 614 cm$^{-1}$ for T$_{FPS}$ $<$ T $<$ T$_{OCO}$, figure \hyperref[fig.PxT.FWHM_Area]{\ref*{fig.PxT.FWHM_Area} (a)}, can be attributed to the differences in the nature of both warming protocols. However, a more detailed study is needed.

Figures \ref{fig.PxT.Centers} and \ref{fig.PxT.FWHM_Area} show that, in relation to the general behavior of the three main structures in the Raman spectrum of LPCMO ($y$ = 0.35), the temperature calibration performed as a function of power is adequate.

\subsection{Raman experiment}

\begin{figure}[H]
    \centering
    \includegraphics[width=\linewidth]{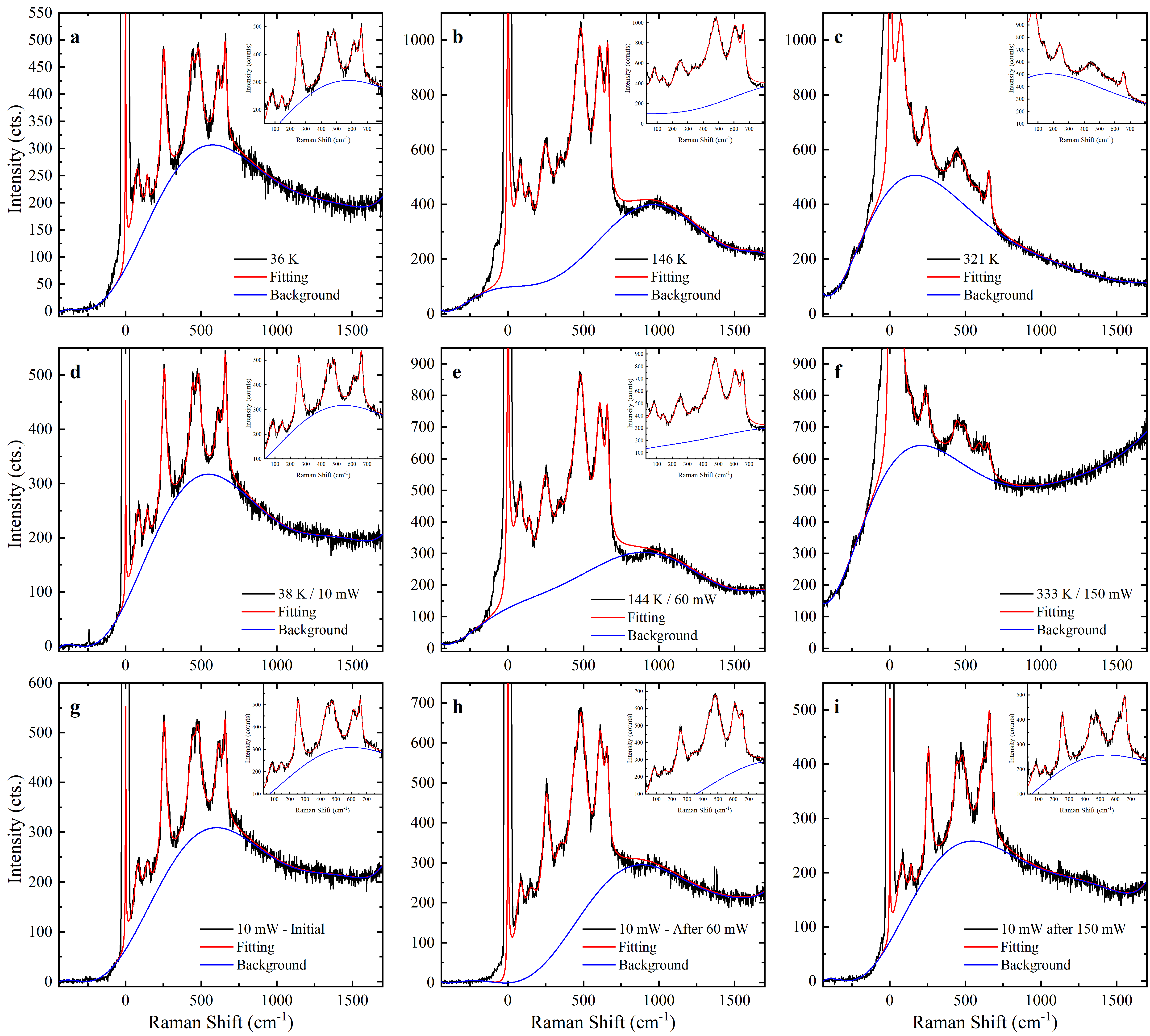}
    \caption{Fitting of the LPCMO Raman spectra for different temperatures, \textbf{a} - \textbf{c}, laser power, \textbf{d} - \textbf{f}, and laser cycles, \textbf{g} - \textbf{i} The blue line it's the adjusted 7$^{th}$ degree polynomial and the red curve it's the sum of the Lorentzian curves and polynomial background for each spectrum shown. In the figures inset it is possible to see the good agreement between the fitting model and data.}
    \label{fig.Raman.Fit}
\end{figure}

The whole frequency range was fitted using a fixed predefined 7$^{th}$ degree polynomial and a group of Lorentzian curves in a Python script. The elastic scattering and the Anti-Stokes contributions were considered part of the background. The fitting quality was judged based on the visual inspection of the model against the data. These can be found in figures \hyperref[fig.Raman.Fit]{\ref*{fig.Raman.Fit} (a) - (i)}.

\subsection{Additional Information}\label{Sup.Mat.Data}

\begin{figure}[H]
    \centering
    \includegraphics[width=\linewidth]{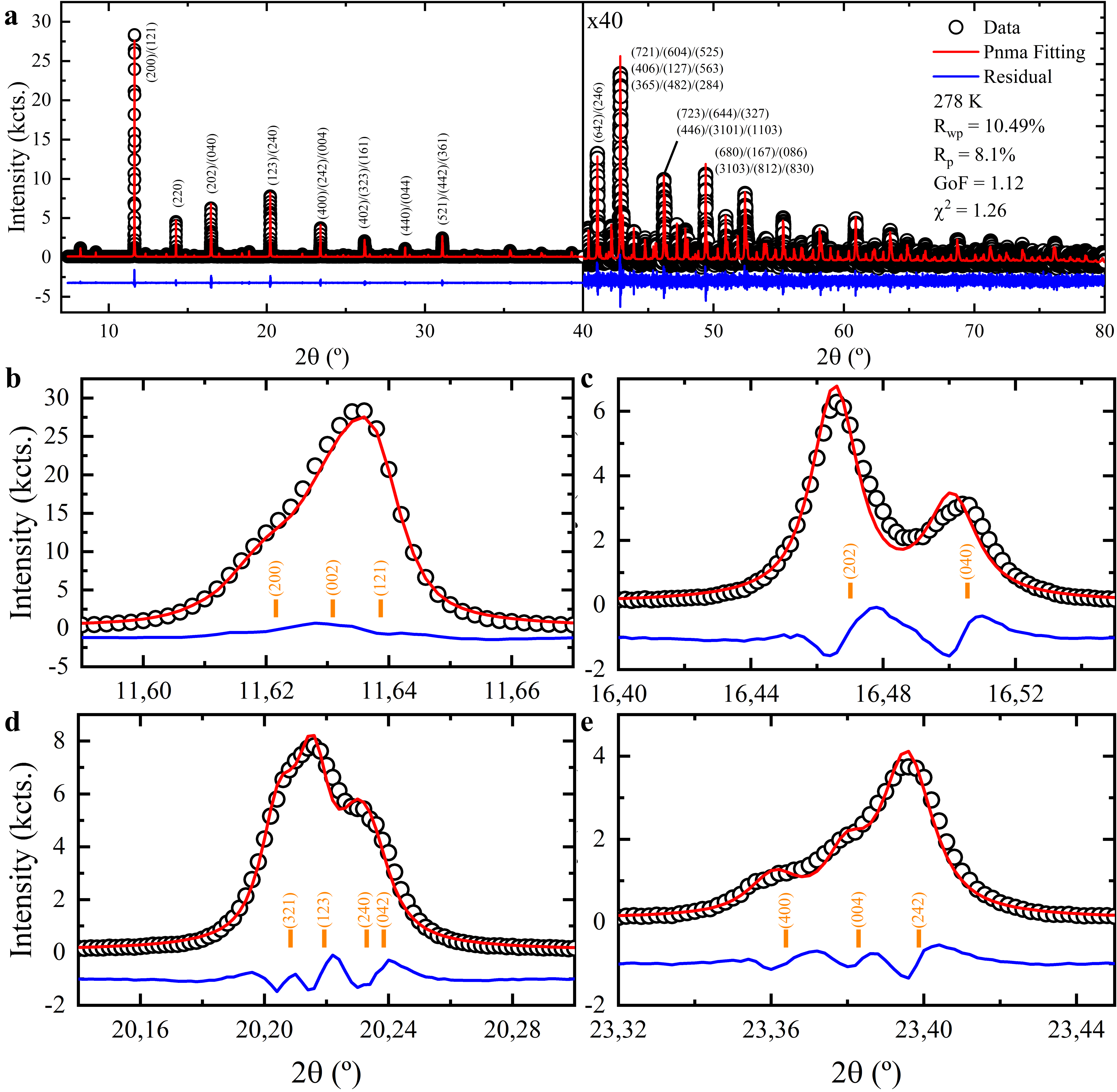}
    \caption{Synchrotron PXRD and Rietveld refinement of LPCMO sample. In \textbf{a} the whole collected angular ranger it's shown together with the Pnma (hkl) indices of the most prominent peaks, as well as in its inset. In figures \textbf{b}-\textbf{e} selected peaks are shown to display the overall good agreement between the data and the fitting. The PXRD pattern show good agreement with the Pnma structure available in the ICSD database \cite{Collado2003} and no extra/lacking peaks were observed. The simplest model was used for the refinement, \textit{i.e.}, only isotropic grain-size, $\mu_{strain}$ and Debye-Waller thermal displacement terms.}
    \label{fig.XRD.ref}
\end{figure}

\begin{table}[H]
\center
\begin{tabular}{cccccc}
\multicolumn{6}{c}{Rietveld Refinement Parameters}\\
\midrule
\multicolumn{2}{c}{R$_{wp}$ (\%)} &  \multicolumn{2}{c}{R$_{p}$ (\%)}  & \multicolumn{2}{c}{$\chi^2$} \\
\multicolumn{2}{c}{10.49} & \multicolumn{2}{c}{8.1}  & \multicolumn{2}{c}{1.26}\\
\midrule
\multicolumn{1}{c}{a (\AA)} & \multicolumn{2}{c}{b (\AA)} & \multicolumn{1}{c}{c (\AA)} & \multicolumn{2}{c}{Volume (\AA$^3$)} \\
\multicolumn{1}{c}{5.44318(1)} & \multicolumn{2}{c}{7.67835(2)} & \multicolumn{1}{c}{5.43884(1)} & \multicolumn{2}{c}{227.3145(6)}\\
\midrule
Site & x & y & z & Frac. & U$_{iso}$\\
\midrule
La & 0.02507(6) &  0.25000 & -0.0049(1) & 0.275 & 0.00565(6)\\
Pr & 0.02507(6) &  0.25000 & -0.0049(1) & 0.35 & 0.00565(6)\\
Ca & 0.02507(6) &  0.25000 & -0.0049(1) & 0.375 & 0.00565(6)\\
Mn & 0.0000 & 0.0000 & 0.5000 & 1.000 & 0.00273(9)\\
O1 & -0.0101(6) & 0.250000 & 0.4388(8) & 1.000 & 0.007(1)\\
O2 & 0.7152(7) & -0.0367(4) & 0.2847(7) & 1.000 &  0.0153(8)\\
\midrule
\bottomrule
\end{tabular}
\caption{Rietveld refinement statistical parameters and Pnma refined unit cell \cite{Collado2003} obtained in the GSAS-II software \cite{GSAS}.}
\label{tab.XRD}
\end{table}

\begin{figure}[H]
    \centering
    \includegraphics[width=\linewidth]{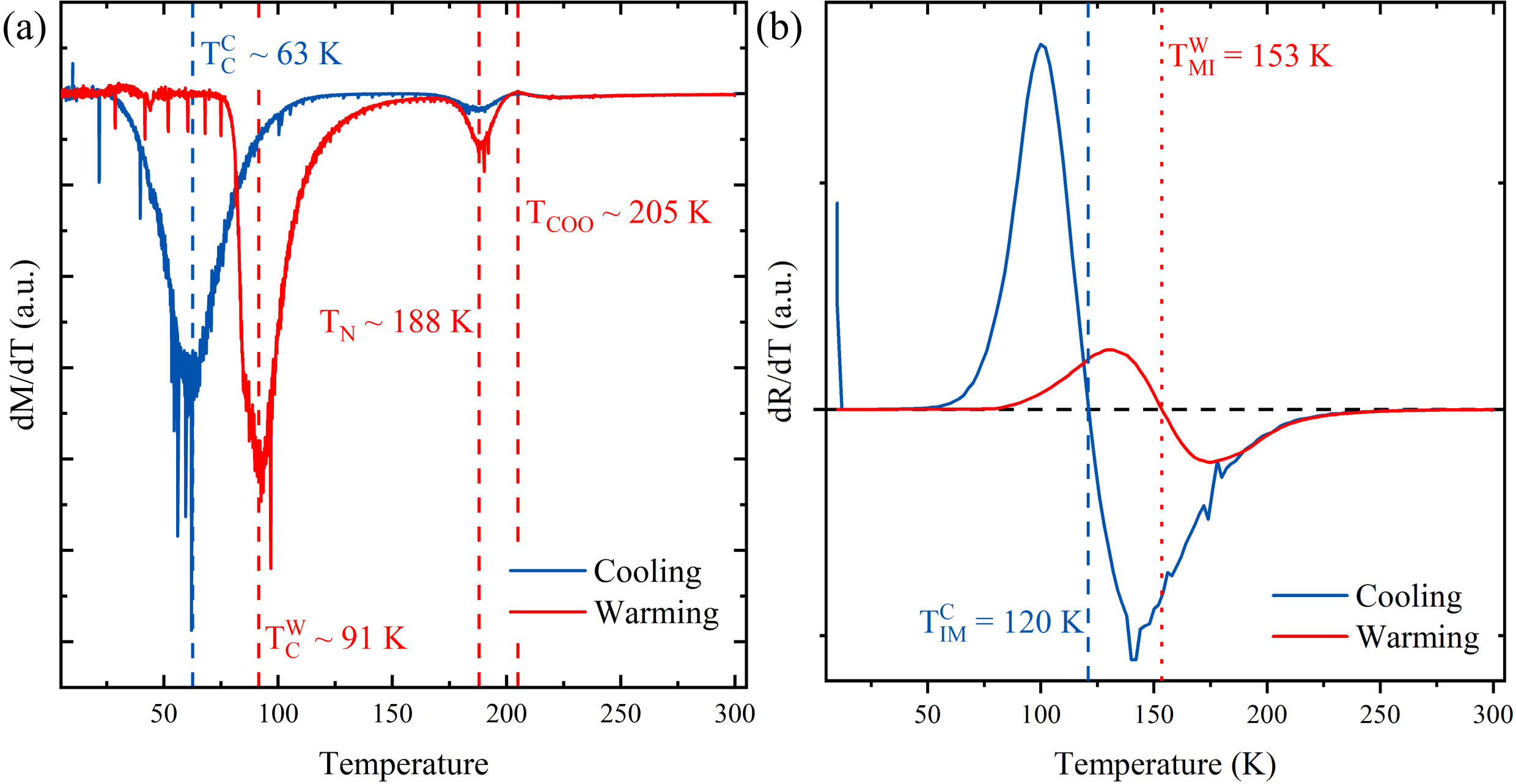}
    \caption{First derivative of the magnetic moment and resistance with respect to the temperature for the cooling (blue) and warming (red) curves, \textbf{a} and \textbf{b}, respectively. The vertical lines indicates the different magnetic and electric transitions observed.}
    \label{fig.dMdT}
\end{figure}

\end{document}